\documentstyle[preprint,aps]{revtex}            
\begin{document}
\draft
\preprint{MKPH-T-95-22}

\title{Coherent $\pi^0$ and $\eta$ Photoproduction on the Deuteron}

\author{S. S. Kamalov\cite{Sabit} and L. Tiator}
\address{Institut f\"ur Kernphysik, Universit\"at Mainz, 55099 Mainz, Germany}

\author{C. Bennhold}
\address{Center for Nuclear Studies, Department of Physics,
The George Washington University,\\
Washington, D.C., 20052, USA}

\maketitle
\begin{abstract}
Coherent pion photoproduction, 
$d(\gamma,\pi^0)d$, 
is calculated in a coupled channels approach
in momentum space using an elementary production operator
that includes Born terms, vector meson exchange and $M1/E2$ $\Delta$-resonance
excitation. The final state interaction is treated in the KMT  
multiple scattering approach that describes $\pi d$ elastic scattering well.
Differential cross sections are found to agree well with existing
($\gamma, \pi^0$) data. The polarization observables $\Sigma , iT_{11}$
and the tensor analyzing powers are predicted for forthcoming experiments. 
A number of transparent relationships between polarization observables 
and the elementary production operator is derived. Finally, calculations 
are performed for coherent eta photoproduction, $d (\gamma, \eta)d$, in the 
threshold region which agree well with a very recent Mainz experiment.
\end{abstract}


\section{INTRODUCTION}
For a long time, coherent pion photoproduction on the deuteron
has been studied as a source of information on $\pi^0$ photoproduction
off the neutron. In the
beginning, the impulse approximation was the main instrument
for this undertaking. However, in contrast to charged pion photoproduction
where the largest contribution comes from the well-known Kroll-Ruderman
amplitude, the isoscalar $\pi^0$ amplitude on the deuteron target
is drastically reduced
due to an accidental cancellation between leading terms.
Therefore, the contributions from pion rescattering
or final state interaction (FSI) become extremely important.
This dramatic effect from pion rescattering with charge exchange
contributions was first found by  Koch and  Woloshyn~\cite{Koch}
and then verified by Bosted and Laget~\cite{Bosted} in studies of coherent
pion photoproduction on  the deuteron in the threshold region.

With increasing photon energy the contribution from the direct
term (without rescattering) becomes very large due to the dominance of the
$\Delta$-resonance contribution and
finally dominates the cross section. In the $\Delta$-resonance
region the main mechanism of FSI  is
elastic pion scattering; the contributions from the
charge exchange reactions become small.
In recent years, the three-body approach has been employed for the
description of FSI. By solving the Faddeev
equations it is possible to take into account the coupling with the
break-up and pion absorption channels.
In this framework considerable progress has been achieved in the
description of pion-deuteron elastic scattering
(see, for example, the review of Ref.~\cite{Garc90}).
Very recently, Wilhelm and Arenh\"ovel \cite{Wil94,Wil95} used an approach
of $NN-N\Delta$
coupled channels for describing deuteron photodisintegration,
$\gamma\,d\rightarrow\,pn$ and coherent pion photoproduction,
\begin{equation}
\gamma + d \rightarrow \pi^0 + d.
\end{equation}

In another approach, recently developed for reaction (1) by
Garcilazo and de Guerra~\cite{Garc95}, relativistic
Feynman diagrams have been evaluated. Using the spectator-on-mass-shell 
prescription, differential  cross sections and polarization
observables have been calculated in the energy region from threshold  
up to 1 GeV.
Blaazer~\cite{blaazer95}, on the other hand, used a Faddeev
approach to the $\pi NN$ system to describe pion rescattering in the
final state.

This paper is the continuation of our previous work which was 
devoted to pion scattering~\cite{KTB95,KTB93} and pion 
photoproduction~\cite{KTB92,KTB91} on very light nuclei. Using
a microscopic approach based on 
the KMT multiple scattering approach~\cite{KMT}
in momentum space we have achieved  a good description of
pion scattering on deuteron and $^3$He and pion photoproduction on $^3$He.

In Part I~\cite{KTB95}, we demonstrated that
a microscopic description of elastic $\pi d$ scattering
in the framework of the multiple scattering theory is able to describe
the differential cross section and polarization observables equally well
as in the Faddeev approach. Furthermore, our multiple scattering approach
proved to be very transparent, simplifying a comparison of the
pion interaction with heavier nuclei. We believe the same rational
holds in the case of coherent  $\pi^0$ photoproduction.

Very recently, interest in the physics with eta mesons has grown
significantly. This relatively new field of theoretical and 
experimental studies is promising to be very useful to obtain a more global
picture of the meson-nuclear interaction. On the other hand, these studies 
could give important information on the excitation of nucleon resonances
in nuclei (like the $S_{11}$(1535) and the $D_{13}$(1525)). 
Eta photoproduction on the deuteron is of special interest since,
due to isospin selection rules a measurement of this process should
allow us to completely determine the proton and neutron electromagnetic
amplitudes and to separate the isospin components of the nucleon resonances.

Unfortunately, despite its long history, the experimental data
available for the meson photoproduction processes on the deuteron are
not of as high a quality as the ones for, i.e., elastic pion scattering.
However, with the new accelerators at Mainz, Bonn, Saskatoon and,
in the near future, CEBAF we hope that such data will
be available soon.

The basic ingredients of our formalism based on the multiple scattering
approach and coupled-channels method are given in Section 2.
In Section 3 we present the main expressions for the differential cross 
section and polarization observables. Our results for coherent $\pi^0$
photoproduction are discussed in Section 4. 
In Section 5, we will complete our analysis with the investigation of coherent
eta photoproduction on the deuteron.
A summary and conclusions are given in Section 6.

\section{BASIC INGREDIENTS OF THE THEORY}

To obtain the amplitude of pion photoproduction on nuclei one may start, 
as in pion scattering, with the amplitude of the elementary process and 
the use of multiple scattering theory~\cite{KMT,Watson}. An extension of 
the KMT approach~\cite{KMT}
to the case of nuclear pion photoproduction leads to the following 
expression for the $T$-matrix:
\begin{equation}
 T_{\gamma}(E)=\sum_{j=1}^A \tau_j^{\gamma} + 
 \sum_{i\neq j}^A \sum_{j=1}^A \tau_i\,G_0(E)\,\tau_j^{\gamma} +
 \sum_{k\neq i}^A \sum_{j\neq i}^A \sum_{j=1}^A \tau_k\,G_0(E)\,\tau_i\,
 G_0(E)\,\tau_j^{\gamma} + ...\,,
\end{equation}
where $\tau_i$ and $\tau_j^{\gamma}$ describe the pion scattering and pion
photoproduction on bound nucleons, respectively, and $G_0(E)$ is the
Green's function for a free pion-nuclear system. Together with the equations 
for the pion-nuclear scattering T-matrix (see Part I), Eq. (2)
can be rewritten as a system of integral equations:  
\begin{mathletters}
\begin{equation}
T_{\gamma}(E)=U_{\gamma}(E)+T'(E)G_0(E)U_{\gamma}(E)\,,\\
\end{equation}
\begin{equation}
T'(E)=U'(E)+U'(E)G_0(E)T'(E)\,,\\
\end{equation}
\end{mathletters}
where
\begin{equation}
T'(E)=\frac{A-1}{A}T(E),\quad U'(E)=\frac{A-1}{A}\sum_{j=1}^A \tau_j(E) 
,\,\mbox{and} \quad U_{\gamma}(E)=\sum_{j=1}^A \tau_j^{\gamma}(E)\,.
\end{equation}

For a consistent derivation of the pion photoproduction amplitude from 
Eq. (3) we will use the same approximations as in pion 
scattering. First in the spectral decomposition of the Green's function only
the deuteron ground state is retained, {\it neglecting the contributions from 
the coupling with the break-up channels}. As we have seen in Part I, 
this approximation is a reasonable starting point to study pion-nuclear
scattering. However, as we will see below, near threshold for isoscalar 
nuclei like the deuteron the ground state contribution is very small
and the coupling to the break-up channels becomes important.
Second we use the free elementary $t$-matrix, $\tau^{\gamma}(E)\approx
t^{\gamma}(\omega)$. The last assumption is called {\it modified impulse 
approximation} where the connection between pion-nucleon ($\omega$) and 
pion-nuclear ($E$) energies is the same as in elastic pion scattering 
(three-body choice):      
\begin{equation}
\omega  =  E+m_\pi+M_N- \left[ (m_\pi+M_N)^2+
\frac{(\vec{q}+\vec{k})^2}{16} \right]^{\frac{1}{2}}-
\left[M_N^2+\frac{(\vec{q}+\vec{k})^2}{16}\right]^{\frac{1}{2}}\,,
\end{equation}
where $m_{\pi}$ and $M_N$ are the pion and nucleon masses, $\vec{k}$ and
$\vec{q}$ are the photon and pion momenta, respectively.

Finally, using these two approximations for the nuclear pion photoproduction 
amplitude we get an expression in the {\it distorted wave impulse 
approximation} (DWIA). In momentum space it can be presented as  
\begin{equation}
 F_{M_f M_i}^{(\lambda)}(\vec{q},\vec{k}) = V_{M_f M_i}^{(\lambda)}
(\vec{q},\vec{k})-\frac{1}{(2\pi)^2} \sum_{M'_f} \int \frac{d\vec{q}\,'}
{{\cal{M}}(q')}\frac{ F'_{M_f M'_f}(\vec{q},\vec{q}\,')
 V_{M'_f M_i}^{(\lambda)}(\vec{q}\,',\vec{k})}{E(q) - E(q') + i\epsilon} 
\, \, ,
\end{equation}
where the amplitudes $F_{M_f M_i}^{(\lambda)}$ and $F'_{M_f M'_f}$ are
connected with  the $T_{\gamma}(E)$ and $T'(E)$-matrixes in Eq. (3)
via relations
\begin{mathletters}
\begin{eqnarray}
F_{M_f M_i}^{(\lambda)}(\vec{q},\vec{k}) =
-\frac{\sqrt{{\cal M}(q){\cal M}(k)}}{2\pi}<\pi(\vec{q}),f 
\mid T_{\gamma}(E) \mid i,\gamma (\vec{k}\lambda)>\,,
\end{eqnarray}
\begin{eqnarray}
F'_{M_f M'_f}(\vec{q},\vec{q}\,')=
-\frac{\sqrt{{\cal M}(q){\cal M}(q\,')}}{2\pi}<\pi(\vec{q}),f 
\mid T'(E) \mid f',\pi (\vec{q}\,')>\,.
\end{eqnarray}
\end{mathletters}
In Eqs. (6-7) $\lambda=\pm 1$ is the photon polarization, 
$\mid i>,\,\mid f>$ and $\mid f'>$ are the deuteron ground state which 
differs only by projection of the deuteron's spin, $M_i,\,M_f$ and
$M'_f$. The relativistic pion-nuclear and photon-nuclear reduced masses
are given by ${\cal M}(q)=E_{\pi}(q)E_{A}(q)/E(q)$ and
${\cal M}(k)=kE_{A}(k)/E(k)$, where $E=E(q)=E(k)$ is the 
pion-nuclear (or photon-nuclear) total energy in the pion-nuclear ($\pi A$) 
c.m. system.

Note that the main difference compared 
to the standard DWIA approach (which is not 
appropriate for the deuteron) is the factor $(A-1)/A$ in the $T'$-matrix.
This factor avoids double counting of pion rescattering on one and the 
same nucleon. Such effects are already included in the elementary amplitude.  
Another difference is that Eq. (6) includes not only coherent (non spin-flip) 
pion rescattering but also incoherent spin-flip transitions which change the
projection of the spin of the deuteron.

In the framework of the {\it plane wave impulse approximation} (PWIA) the 
pion photoproduction amplitude is equal to the first term of Eq. (6)
which is expressed in terms of the elementary pion-nucleon photoproduction 
$t^{\gamma}(\omega)$-matrix or amplitude $f^{(\lambda)}$: 
\begin{mathletters}
\begin{equation}
 V_{M_f M_i}^{(\lambda)}(\vec{q},\vec{k})=
-\frac{\sqrt{{\cal M}(q){\cal M}(k)}}{2\pi}<\pi (\vec{q}),f \mid
\sum^A_{j=1}\,t^{\gamma}_j(\omega)\,\mid i,\gamma (\vec{k}\lambda)>\,,
\end{equation}
\begin{equation}
<\vec{q},\vec{p}\,'\mid t^{\gamma}(\omega)\mid \vec{p},\vec{k}\lambda>=
-\frac{2\pi}{\sqrt{\mu (q,p')\mu (k,p)}}\,f^{(\lambda)}
(\omega,\theta_{\pi}^*)\,
\delta(\vec{p}\,'+\vec{q}-\vec{p}-\vec{k})\,,
\end{equation}
\end{mathletters}
where relativistic pion-nuclear and photon-nuclear reduced masses
are  ${\mu}(q,p')=E_{\pi}(q)E_{N}(p')/\omega$ and $\mu(k,p)=kE_{N}(p)/\omega$,
$\vec{p}$ and $\vec{p}\,'$ are the nucleon momenta in the initial and final 
states (in $\pi A$ c.m. system), respectively. 

The relativistically invariant amplitude  $f^{(\lambda)}$, in the 
pion-nucleon ($\pi N$) c.m. system has the standard CGLN form~\cite{CGLN}
\begin{equation}
f^{(\lambda)}(\omega,\theta_{\pi}^*)=
i F_{1}\vec{\sigma}\cdot\vec{\epsilon}_{\lambda}+
  F_{2}\vec{\sigma}\cdot\hat{\vec q}\,\,\vec{\sigma}\cdot
[\hat{\vec{k}}\times\vec{\epsilon}_{\lambda}]+
iF_{3}\vec{\sigma}\cdot\hat{\vec k}\,\,\hat{\vec q}\cdot
\vec{\epsilon}_{\lambda}+i F_{4}\vec{\sigma}\cdot\hat{\vec q}\,\,
\hat{\vec{q}}\cdot\vec{\epsilon}_{\lambda}\,,
\end{equation}
where $\hat{\vec q}$ and $\hat{\vec k}$ are the unit vectors for the
pion and photon momenta in the $\pi N$ c.m. system, $\theta^*_{\pi}$ is the 
pion angle in the same system. All these variables can be expressed 
in an arbitrary frame via the standard Lorentz transformation 
(see Ref.~\cite{KTB91}).

In nuclear applications, it is convenient to divide the amplitude in Eq. (9)
into non spin-flip ($f_2$) and spin-flip ($f_1,\,f_3,\, f_4$) amplitudes
\begin{equation}
f^{(\lambda)}=
i f_{1}\vec{\sigma}\cdot\vec{\epsilon}_{\lambda}+
  f_{2}[\hat{\vec{k}}\times\hat{\vec q}]\cdot{\epsilon}_{\lambda}+
if_{3}\vec{\sigma}\cdot\hat{\vec k}\,\,\hat{\vec q}\cdot
\vec{\epsilon}_{\lambda}+i f_{4}\vec{\sigma}\cdot\hat{\vec q}\,\,
\hat{\vec{q}}\cdot\vec{\epsilon}_{\lambda}\,.
\end{equation}
These are connected to the standard CGLN amplitudes via
\begin{equation}
f_1=F_1-F_2\cos{\theta^*_{\pi}},\quad f_2=F_2, \quad f_3=F_3+F_2,
\quad f_4=F_4\,.
\end{equation}

In terms of the lowest $s$- and $p$-wave multipoles,
the amplitudes $f_i$ can be expressed as
\begin{mathletters}
\begin{equation}
 f_1=E_{0+}+\cos{\theta_{\pi}^*}(3\,E_{1+}+M_{1+}-M_{1-})\,,
\end{equation}
\begin{equation}
 f_2=2\,M_{1+}+M_{1-}\,,
\end{equation}
\begin{equation}
 f_3=3\,E_{1+}-M_{1+}+M_{1-}\,,
\end{equation}
\begin{equation}
 f_4=0\,.
\end{equation}
\end{mathletters}

In our calculation,  we use the unitary version of the Blomqvist-Laget 
amplitude~\cite{BL2,BL3} for the elementary amplitude
which contains contributions from Born and $\omega$-exchange terms.
The dominant $\Delta$-resonance contribution is described by the real 
and imaginary parts of the resonant $M_{1+}^{\Delta}(\omega)$ and 
$E_{1+}^{\Delta}(\omega)$ multipoles that are added to the background terms 
in a unitarized way. In the framework of this model the elementary process is 
well described for photon energies up to  
$E_{\gamma}$=500 MeV.

The explicit expression for the full elementary amplitude is given 
in Ref.~\cite{KTB91}. We only remind that in accordance with the
Blomqvist-Laget approach it is obtained after a non-relativistic reduction 
(up to order $(\vec{p}/M_N)^2$) of the relativistic and gauge invariant
amplitude. One of the advantages of this model is that the final expression
for the elementary amplitude is well appropriate for nuclear calculations
with non-relativistic wave functions. However, due to the reduction procedure
the gauge invariance is lost. The neglected terms that could
restore the gauge invariance are of the order of $(\vec{p}/M_N)^3$
(consistent with our approximation) and their influence is small.   

Finally, we mention the so-called {\em factorization approximation}
in the treatment of nucleon Fermi motion. We have developed the
formalism for the exact treatment of the nucleon momentum
dependence in the elementary amplitude using a multidimensional integration
in Eq. (7). However, similar to our previous studies~\cite{TW,EGK} 
the substitution 
\begin{equation}
\vec{p} \rightarrow \vec{p}_{eff.}=- \frac{\vec{k}}{A}-
\frac{A - 1}{2A} (\vec{k} - \vec{q}) =-\frac{\vec{k}}{2}
 - \frac{1}{4} \vec{Q}
\end{equation}
works very well quantitatively and provides an excellent approximation.
This result is based on the fact that in case of a 
Gauss wave function, which roughly reproduces the dominant S-wave part of the 
deuteron ground state, the replacement (13) gives an exact treatment for the 
linear $\vec{p}/2M$ terms in the elementary amplitude.   
 
\section{POLARIZATION OBSERVABLES}

Polarization observables have the promise of opening a new field in the
electromagnetic production of pions from protons and nuclear targets.  Since
many of these observables contain interference terms, small but important
amplitudes can be investigated in a unique way.

To define our polarization observables we choose the system of coordinates
corresponding to the {\it Madison Convention}~\cite{Madison}:
a right-handed coordinate system where the positive $z$-axis is along 
the photon beam direction and the $y$-axis is along the vector 
$[\vec{k}\times\vec{q}]$.

In this paper, we focus only on single polarization
observables that appear in $\pi^0$ photoproduction with a polarized beam
and an unpolarized target, or an
 unpolarized  beam and a polarized target. In both
cases five polarization observables can be measured:\\
{\it The photon asymmetry}
\begin{equation}
\Sigma=\frac{d\sigma/d\Omega^{\perp} - d\sigma/d\Omega^{(\parallel)}}{d\sigma/
d\Omega^{\perp} + d\sigma/d\Omega^{(\parallel)}}\,,
\end{equation}
where $\perp (\parallel)$ refers to a photon linearly polarized 
perpendicular (parallel) to the reaction plane
and four {\it analyzing powers} (the vector, $iT_{11}$, and three tensor
ones, $T_{20}, T_{21}, T_{22}$)
\begin{equation}
T_{k\kappa}=\frac{Tr(F\tau_{k\kappa}F^{\dag})}{Tr(F\,F^{\dag})}\,.
\end{equation}
In Eq. (15), the spherical tensor operator $\tau_{k\kappa}$ is defined as
\begin{equation}
 <J\,M'\mid \tau_{k\kappa} \mid J\,M> = {\hat J}{\hat k} (-1)^{J+M'}
\left({\,J\;\;\;\;\;\;\;J\;\;\;\;k}
\atop{\,M\;\;-M'\;\;\kappa}\right)
\end{equation}
with a Wigner $3 j$ symbol
and $\hat{J} = \sqrt{2 J +1}$.

From the partial wave decomposition of the electromagnetic vector
potential$~\cite{Watson}$ 
it follows that the pion photoproduction matrix elements,
$F_{M_f M_i}^{(\lambda)}$, can be divided into magnetic and
electric components which are both independent of the photon
polarization $\lambda=\pm 1$:
\begin{equation}
F_{M_f M_i}^{(\lambda)}\,=\, F_{M_f M_i}^{(M)}\,+\,\lambda\,
F_{M_f M_i}^{(E)}\,.
\end{equation}
Similar to the formalism in pion scattering, 
one can present the photoproduction matrix in the following form 
\begin{equation}
F_{M_f M_i}^{(\lambda)} = \left( \begin{array}{ccc}
F_{++}^{(\lambda)} & F_{0+}^{(\lambda)} & F_{-+}^{(\lambda)} \\
F_{+0}^{(\lambda)} & F_{00}^{(\lambda)} & F_{-0}^{(\lambda)} \\
F_{+-}^{(\lambda)} & F_{0-}^{(\lambda)} & F_{--}^{(\lambda)} \\
\end{array} \right) =
\left( \begin{array}{ccc}
\;{\cal A}_M    &\,{\cal B}_M &\,{\cal C}_M \\
 {\cal D}_M &\,{\cal E}_M &\,{\rm-}{\cal D}_M \\
\;{\cal C}_M    &\!{\rm-}{\cal B}_M &\;{\cal A}_M \\
\end{array} \right) +\lambda
\left( \begin{array}{ccc}
\;{\cal A}_E    &\;{\cal B}_E &\,{\cal C}_E \\
\;{\cal D}_E & 0 &\,{\cal D}_E \\
{-\cal C}_E   &\;\;{\cal B}_E & {\rm-}{\cal A}_E \\
\end{array} \right)\,,
\end{equation}
where the signs $+,0,-$ correspond to the deuteron spin projections
$M_{i(f)}=+1,0,-1$, respectively.  Note that
the structure of the magnetic part
of matrix (18) is similar to the {\it Robson} matrix~\cite{Robson} 
which was discussed in the case of elastic pion scattering~\cite{KTB95}. 
The main difference appears in the electric part, whose contribution is 
much smaller. Therefore, the behavior of the differential cross section
and the single polarization observables is mainly determined
by the five magnetic amplitudes. We point out that the ${\cal C}_E$
amplitude vanishes in PWIA; it has contributions due only to pion rescattering.

Calculating the expectation values of the operator $\tau_{k\kappa}$ 
for $k=\kappa=0$, the differential cross section and photon asymmetry
can be written as
\begin{mathletters}
\begin{equation}
\frac{d\sigma}{d\Omega}=\frac{d\sigma_M}{d\Omega}+
\frac{d\sigma_E}{d\Omega}\,,
\end{equation}
\begin{equation}
\frac{d\sigma_M}{d\Omega}=\frac{2q}{3k}(\mid {\cal A}_M \mid^2+
\mid {\cal B}_M \mid^2 + \mid {\cal C}_M \mid^2+
\mid {\cal D}_M \mid^2 +\frac{1}{2} \mid {\cal E}_M \mid^2)\,,
\end{equation}
\begin{equation}
\frac{d\sigma_E}{d\Omega}=\frac{2q}{3k}(\mid {\cal A}_E \mid^2+
\mid {\cal B}_E \mid^2 + \mid {\cal C}_E \mid^2+\mid {\cal D}_E \mid^2)\,,
\end{equation}
\begin{equation}
\Sigma=\frac{d\sigma_M/d\Omega - d\sigma_E/d\Omega}{d\sigma_M/
d\Omega + d\sigma_E/d\Omega}\,.
\end{equation}
\end{mathletters}

Similarly, the vector and tensor analyzing powers can be divided into
magnetic and electric parts
\begin{equation}
  T_{k\kappa}=T_{k\kappa}^{(M)}+T_{k\kappa}^{(E)}\,.
\end{equation}
The expression for $T_{k\kappa}^{(M)}$ in terms of
the ${\cal A}_M,...,{\cal E}_M$
amplitudes are the same as in pion scattering:
\begin{mathletters}
\begin{equation}
iT_{11}^{(M)}=\sqrt{\frac{2}{3}}\,Im\,[\,{\cal D}_M^*({\cal A}_M-
{\cal C}_M)-{\cal B}_M^*{\cal E}_M)\,]/a (\theta_{\pi})\,,
\end{equation}
\begin{equation}
T_{20}^{(M)}=\frac{\sqrt{2}}{3}(\mid {\cal A}_M \mid^2+
\,\mid {\cal B}_M \mid^2 + \mid {\cal C}_M \mid^2 -
2 \mid {\cal D}_M \mid^2 -\mid {\cal E}_M \mid^2)/a (\theta_{\pi})\,,
\end{equation}
\begin{equation}
T_{21}^{(M)}=-\sqrt{\frac{2}{3}}\,Re\,[\,{\cal D}_M^*\,({\cal A}_M-
{\cal C}_M)+{\cal B}_M^*{\cal E}_M)\,]/a (\theta_{\pi})\,,
\end{equation}
\begin{equation}
T_{22}^{(M)}=\frac{1}{\sqrt{3}}\,[\,2\,Re\,(\,{\cal A}_M^*\,{\cal C}_M)-
\mid {\cal B}_M \mid^2]/a (\theta_{\pi})\,,
\end{equation}
\begin{equation}
a (\theta_{\pi}) = \frac{k}{q} \frac{d\sigma}{d\Omega}\,.
\end{equation}
\end{mathletters}

For the electric part of the analyzing powers we obtain
\begin{mathletters}
\begin{equation}
iT_{11}^{(E)}=\sqrt{\frac{2}{3}}\,Im\,[\,{\cal D}_E^*({\cal A}_E +
{\cal C}_E)\,]/a (\theta_{\pi})\,,
\end{equation}

\begin{equation}
T_{20}^{(E)}=\frac{\sqrt{2}}{3}(\mid {\cal A}_E \mid^2+
\,\mid {\cal B}_E \mid^2 + \mid {\cal C}_E \mid^2 -
2 \mid {\cal D}_E \mid^2)/a (\theta_{\pi})\,,
\end{equation}
\begin{equation}
T_{21}^{(E)}=-\sqrt{\frac{2}{3}}\,Re\,[\,{\cal D}_E^*\,({\cal A}_E +
{\cal C}_E)\,]/a (\theta_{\pi})\,,
\end{equation}
\begin{equation}
T_{22}^{(E)}=-\frac{1}{\sqrt{3}}\,[\,2\,Re\,(\,{\cal A}_E^*\,{\cal C}_E)-
\mid {\cal B}_E \mid^2]/a (\theta_{\pi})\,.
\end{equation}
\end{mathletters}

Finally, we present expressions for the ${\cal A},\,{\cal B},\,etc.$
amplitudes obtained: 1) in the plane wave impulse approximation,
2) by taking into account  only contributions from the $S$-component and its 
interference with the $D$-component of the deuteron wave function.
Then the Robson amplitudes can be expressed solely through the elementary
amplitudes $f_1,...,f_4$ from Eq. (10) and the radial integrals $R_{l'l}$ 
(introduced in Part I):
\begin{equation}
R_{l'l}^{(L)}(Q)=\int_{}^{}d r\,U_{l'}(r)\,j_L(\frac{Qr}{2})\,U_l(r),
\end{equation}
where $U_l$ is the deuteron radial wave function  
$j_L(z)$ is the spherical Bessel function, and $\{l, l'\} = 0,2$.
This yields
\begin{mathletters}
\begin{equation}
{\cal A}_M\cong\sqrt{2}f_2\sin{\theta_{\pi}^*}
\left[R_{00}^{(0)} -\frac{1}{2\sqrt{2}}(1+3\cos{2\Theta})\,
R_{02}^{(2)}\right]\,W_A\,,
\end{equation}
\begin{equation}
{\cal B}_M\cong \left(f_1\,R_S+\frac{3}{\sqrt{2}}f_2\sin{\theta_{\pi}^*}
\sin{2\Theta}\,R_{02}^{(2)}\right)\,W_A\,,
\end{equation}
\begin{equation}
{\cal C}_M\cong -\frac{3}{2}f_2\sin{\theta_{\pi}^*}\left(1 - \cos{2\Theta}
\right)\,R_{02}^{(2)}\,W_A\,,
\end{equation}
\begin{equation}
{\cal D}_M\cong -\left(f_1\,R_S-\frac{3}{\sqrt{2}}f_2\sin{\theta_{\pi}^*}
\sin{2\Theta}\,R_{02}^{(2)}\right)\,W_A\,,
\end{equation}
\begin{equation}
{\cal E}_M\cong \sqrt{2}f_2\sin{\theta_{\pi}^*}
\left[R_{00}^{(0)} +\frac{1}{\sqrt{2}}(1+3\cos{2\Theta})\,
R_{02}^{(2)}\right]\,W_A\,,
\end{equation}
\begin{equation}
{\cal A}_E\cong -\sqrt{2}\sin{\theta_{\pi}^*}\left[(f_3+f_4\cos{\theta_{\pi}^*})
R_S+\frac{3}{\sqrt{2}}f_Q\cos{\Theta}\,R_{02}^{(2)}\right]\,W_A\,,
\end{equation}
\begin{equation}
{\cal B}_E={\cal D}_E\cong -\left[(f_1+f_4\sin^2{\theta_{\pi}^*})R_S-
\frac{3}{\sqrt{2}}f_Q\sin{\theta_{\pi}^*}\sin{\Theta}\,R_{02}^{(2)}
\right]\,W_A\,,
\end{equation}
\end{mathletters}
where we have used the following notations:
\begin{mathletters}
\begin{equation}
R_S(Q)=R_{00}^{(0)}(Q)+\frac{1}{\sqrt{2}}R_{02}^{(2)}(Q)\,,
\end{equation}
\begin{equation}
f_Q=\frac{q}{Q}(f_1+f_3\cos{\theta_{\pi}^*}+f_4)-
\frac{k}{Q}(f_3+f_4\cos{\theta_{\pi}^*})\,,
\end{equation}
\begin{equation}
\cos{\Theta}=\frac{1}{Q}(k-q\cos{\theta_{\pi}})\,.
\end{equation}
\end{mathletters}
In Eqs. (24-25) $\theta_{\pi}^*$ is the pion angle in $\pi N$
c.m. system, $\vec{Q}=\vec{k}-\vec{q}$ is the nuclear momentum transfer.
The kinematical factor $W_A$ can easily be obtained from Eq. (8).
 
As we will see below, the expressions (24) are very useful for a
qualitative as well as a quantitative understanding of the polarization 
observables. The complete expressions (including complete 
contributions from the deuteron $D$-state component) are given in the 
Appendix.  

\section{ RESULTS AND DISCUSSION}
 
\subsection{Differential cross section}
 
We begin our discussion with the results for the differential cross
section. Using our formalism described above we will show that it reproduces 
the well-known results for the coherent $\pi^0$ photoproduction at threshold 
and in the $\Delta$-resonance region.

In general, pion photoproduction near threshold is dominated by
the spin-flip amplitude with the leading multipole $E_{0+}$.
However, neutral pion photoproduction is
comparable to the situation with the scattering length in elastic pion-deuteron
scattering (see Part I): 
the isoscalar part of the s-wave multipole is one order of magnitude 
smaller than the isovector one. Therefore, the  PWIA approach
is not valid at threshold and contributions from pion rescattering with 
charge exchange become very important~\cite{Koch,Bosted}. 
For this two-step process the corresponding contribution to the 
total amplitude can be estimated by using the first iteration 
of Eq. (3)
\begin{equation}
F^{(\lambda)}_{Resc.}=\frac{{\cal M}_{\pi d}}{\mu_{\pi N}}\,W_A\,
< d \mid e^{i{\vec r}\cdot({\vec k}+{\vec q})/2}\frac{e^{iq_0r}}{r}
\sum_{i \neq j}\,f_{\pi N}(i)\,f^{(\lambda)}(j)\,\mid d >\,,
\end{equation}
where ${\cal M}_{\pi d}$ and $\mu_{\pi N}$ are the reduced masses for the
$\pi d$ and $\pi N$ systems, respectively, 
$\mid d>$ is the deuteron ground state 
and  $q_0$ is the on shell momentum of the pion in the intermediate 
state. 
    
In the limit of $q \rightarrow 0$, the elementary amplitudes for $\pi N$
scattering, $f_{\pi N}$, and pion photoproduction, $f^{(\lambda)}$, become
constant and the total amplitude (with direct and rescattering contributions)
can be reduced to the expression~\cite{Laget}
\begin{equation}
F^{(\lambda)}\approx W_A\,R_{00}^{(0)}(Q)
\left[(E_{0+}^{\pi^0p}+E_{0+}^{\pi^0n})+
\frac{{\cal M}_{\pi d}}{\mu_{\pi N}}\left<\frac{1}{r}\right>\,
a_{\pi N}^{\pi^0\pi^+}\,(E_{0+}^{\pi^+n}-E_{0+}^{\pi^-p})\right]
\end{equation}
with the pion-nucleon scattering length 
$a_{\pi N}^{\pi^0\pi^+}=-a_{\pi N}^{\pi^0\pi^-}=\sqrt{2}(a_1-a_3)/3
=0.122/m_{\pi}$ and $<1/r>=0.65/m_{\pi}$. In Eq. (27), the $E_{0+}^{\pi^0 N}$
multipoles for $\pi^0$ photoproduction are much smaller than the Kroll-Ruderman 
multipole $E_{0+}^{\pi^{\pm} N}$, thus the rescattering terms are 
comparatively enhanced.

In Fig. 1 we show our calculations for the total and differential 
cross section in the threshold region.
In this region the contribution of elastic rescattering
(without charge exchange) is very small. This is due to the small size
of the isoscalar part of the $\pi N$ scattering amplitude.  On the other
hand, due to its large isovector part and also due to the large
$E_{0+}^{\pi^{\pm} N}$ multipole, the role of the charge exchange
process becomes very important. 
In the upper figure the reduced cross section is compared with a
recent re-analysis~\cite{Saclay} of an earlier Saclay measurement. The full 
calculations are about 15\% larger than the experimental data.
The same results have recently been obtained by Garcilazo and
Guerra~\cite{Garc95} using a Feynman diagram method with a pure 
pseudovector $\pi NN$ coupling. Calculations by Bosted and
Laget~\cite{Bosted} are about 15\% lower than ours and
agree better with the experiment.

The authors of Ref.~\cite{Garc95} argue that the main reason of 
their disagreement with the experiment could be due to
the contributions from third- and higher-order scattering terms.
However, our calculations (compare dashed and solid curves in Fig. 1) show 
that such contributions usually increase the cross section because of the
attractive nature of the pion-nuclear interaction in the threshold region.

In the lower part of Fig. 1 we show calculations for the differential
cross section at $\Delta E= E_{\gamma}-E_{\gamma}^{thr.}=$ 8 MeV.
Here we see that the interference between direct and rescattering terms 
is strongly angle dependent, leading to an increase in the differential
cross section by about a factor of two at backward angles.

Recent developments in $\pi^0$ photoproduction on the proton 
\cite{Mazzu,Beck} that indicate
large deviations from the old Low Energy Theorem
(LET) predictions suggest that such effects could also be visible in 
coherent photoproduction on the deuteron. In the direct term only the 
isospin (+) amplitude appears which is proportional to the sum of 
the proton and neutron 
amplitudes, $E_{0+}^{(+)}=(E_{0+}(\pi^0p)+E_{0+}(\pi^0n))/2$. While PV
Born terms predict this amplitude to be
 constant in the threshold region with a value
close to the old LET, $E_{0+}^{(+)}=-1.0\cdot10^{-3}/m_\pi$, the new 
experiments on the proton show an energy dependent function with a cusp 
effect at the charged pion threshold. 
Using isospin symmetry, with this new experimental data and the older 
values for $(\gamma,\pi^\pm)$ \cite{Dre92}
we find values for $E_{0+}^{(+)}$ from -1.5 to +1.5 in units of 
$10^{-3}/m_\pi$. 
Predictions from chiral perturbation theory give positive values for this 
amplitude that vary between the neutral and charged thresholds in the
range of $(0.5 - 1.2)\cdot10^{-3}/m_\pi$ \cite{Bern94}.
In Fig. 2, we show the influence of this isospin (+) amplitude on the total
and differential cross section of coherent pion photoproduction on the 
deuteron. Even though the direct term is not dominant in this reaction, a
significant sensitivity
 can be seen because of the interference with the pion rescattering
channel. Comparing our calculations with the
total cross section data in Fig. 2a leads to
a preferred value of $-0.5\cdot10^{-3}/m_\pi$.
As demonstrated in Fig. 2b, a determination of
$E_{0+}^{(+)}$ should be much easier using angular distributions,
 where one can easily distinguish
different values in the forward-backward angular asymmetry.

With increasing photon energy the contributions from $p$-wave 
multipoles become stronger, due mostly to the excitation of the 
$\Delta$-resonance. Fig. 3 illustrates the role
of the individual parts of the elementary photoproduction amplitude: the
Born+$\omega$-exchange terms (dotted curves), the $\Delta$-resonance
contribution (dashed curves) and the total amplitude (solid curves).
The Born and $\omega$-exchange diagrams are essential at forward and backward
angles. In this region only the $f_1$ amplitude remains which contains
the
nonresonant $E_{0+}$ and $M_{1-}$ multipoles as well as the resonant
$M_{1+}^{\Delta}$ and $E_{1+}^{\Delta}$ amplitudes.
Around $\theta_{\pi}$=$90^0$, the $\Delta$-resonance contribution
dominates due to the magnitude
of the coherent contribution from the non spin-flip amplitude $f_2$.  
 
In comparison to the available data the overall agreement is satisfactory.
An exception may be the large disagreement in the forward direction  with
the data from Ref.~\cite{Bouq74} with rather large error bars.
The best description of the experimental measurements
is achieved at $\theta_{\pi}=90^0$, where the $\Delta$ contribution
dominates. Together with our earlier results obtained
for $^{3}He$ and $p$-shell nuclei,
this gives us confidence in the treatment of
$\Delta$ propagation in nuclei. In order to draw further conclusions
more precise experiments are needed, in particular at
extreme angles where the $\Delta$ contribution is weaker.
 
Fig. 4 shows the differential cross sections at $E_{\gamma}=260-400$ MeV.
Here we illustrate the role of the spin-flip contribution
(dash-dotted curves). It only contributes significantly
 at very backward and forward angles.
Around $\theta_{\pi}=90^0$ the main contribution comes from the non spin-flip
transition associated with the elementary amplitude $f_2$ which is
dominated by the $\Delta$-resonance.

The role of the rescattering mechanism can be seen by comparing the simple
PWIA (dashed curves) with our full calculations (solid curves).
Its contribution is biggest in the $\Delta$-resonance region and at 
backward angles. Among the different rescattering mechanisms elastic 
scattering (without charge exchange) dominates in this energy region. 
This is in contrast to the threshold region where
, as we have seen above, charge
exchange rescattering gives the main
contribution.  Above the $\Delta$-resonance
region, the rescattering  contribution becomes smaller
with increasing photon energy.

The role of the $^{3}D_1$ configuration in the deuteron wave function is also
illustrated in Fig. 4 by turning off the nuclear matrix elements with
$L=2$. The $D$-wave contribution becomes manifest at
large angles, $\theta_{\pi}>90^0$, for energies $E_{\gamma}>300$ MeV,
which corresponds to momentum transfers of $Q^2>3.1 fm^{-2}$.

We conclude our analysis of the differential cross section by comparing
with previous calculations. Our
results presented here  are very similar to the ones obtained by
Bosted and Laget~\cite{Bosted}. The main difference
between our and their approach is the treatment of the pion rescattering
contributions. In our formalism the coupling with the break-up
channels was not included in contrast to Ref.~\cite{Bosted}.
Apparently, the corresponding contribution is small in the kinematical range
considered here.

A comparison with the Garcilazo and Guerra~\cite{Garc95} calculations shows 
that in the forward direction their results are generally about 50\% 
larger, except at $E_{\gamma}$=340 MeV. This could indicate 
a difference in the spin-flip part of the elementary amplitude which  
dominates in the forward direction. Below we show that this leads to
a dramatic difference in the vector analyzing power.
The work by Blaazer~\cite{blaazer95} includes only a comparison with the
data at $E_{\gamma}$=300 MeV. Their results are about 20\% above the
measured cross section. Within their Faddeev formalism they find only
small pion rescattering effects.
Compared to the calculations by Wilhelm and Arenh\"ovel \cite{Wil95}
our results are lower and agree better with the data, except for
$\theta_\pi=6^0$. The main reason for this difference is due to pion 
rescattering in our approach which reduces the cross sections considerably
in the region of the $\Delta$-resonance (see Fig. 4).

\subsection{Polarization observables.}

{\it\underline{Photon asymmetry}.} Fig. 5 shows the angular distribution
for the photon asymmetry calculated at $E_{\gamma}=200-400$ MeV
in PWIA (dashed curves) and DWIA (solid curves). The dotted
curves are DWIA calculations without the $D$-state configuration in the
deuteron wave function. Clearly, all calculations are very similar;
thus, the influence of the deuteron $D$-state
and pion rescattering is small. Moreover, our analysis showed that the
contribution of the $l=2$ multipoles in the elementary production amplitude
is also negligible, e.g., we can set $f_4=0$. Thus, using Eq. (24)
the photon asymmetry can be
expressed in a rather simple form using only the elementary amplitudes
$f_1,\,f_2$ and $f_3$:
\begin{equation}
\Sigma^{(S)}(deuteron)\approx\frac{3}{4}\sin^2{\theta_{\pi}}
\frac{\mid f_2 \mid^2-\frac{2}{3}\mid f_3 \mid^2}
{\mid f_1 \mid^2 + \frac{3}{4}\sin^2{\theta_{\pi}}
(\mid f_2 \mid^2 +\frac{2}{3}\mid f_3 \mid^2)}\,.
\end{equation}

This expression is close to the expression for the elementary photon
asymmetry for the process on the nucleon:
\begin{equation}
\Sigma(nucleon)\approx\frac{1}{2}\sin^2{\theta_{\pi}}
\frac{\mid f_2 \mid^2-\mid f_3 \mid^2}
{\mid f_1 \mid^2 + \frac{1}{2}\sin^2{\theta_{\pi}}
(\mid f_2 \mid^2 +\mid f_3 \mid^2)}\,.
\end{equation}
Therefore, the angular distribution of the photon asymmetry for the 
production process on hydrogen and deuteron looks very similar.

{\it\underline{Tensor analyzing powers}.}  Fig. 6 presents our
results for the $T_{20}$ observable obtained in the same way as the
photon asymmetry. However, in contrast to the photon asymmetry, $T_{20}$
is very sensitive to the deuteron $D$-state component (compare
solid and dotted curves). The influence of pion rescattering still
remains small.

In the region of $90^0<\theta_{\pi}<150^0$, where the contribution of
the $D$-state is more important the value for $T_{20}$
can be estimated from Eq. (24) as
\begin{equation}
T_{20}(\sim 130^0)\approx\frac{\mid {\cal A}_M \mid^2 - \mid {\cal E}_M 
\mid^2} {\mid {\cal A}_M \mid^2 +\frac{1}{2} \mid {\cal E}_M \mid^2}
\approx-(1+3\cos{2\Theta})\frac{x}{1+3\,x^2\,\sin^2{2\Theta}}\,,
\end{equation}
where angle $\Theta$ is defined in Eq. (25c) and 
$x=R_{02}^{(2)}(Q)/R_{00}^{(0)}(Q)$ . This ratio describes
the role of the deuteron $D$-state as a function of the
momentum transfer $Q$. It is clear that with increasing $Q$ the ratio $x$
also increases and $\mid T_{20} \mid$ reaches its maximum value.

The contribution of the deuteron $S$-state 
dominates at forward and backward angles. In this region 
${\cal B}_M={\cal D}_M=-{\cal B}_E=-{\cal D}_E$ and other amplitudes
are zero. Therefore, independent of the photon energy we have
\begin{equation}
T_{20}(0^0)\approx T_{20}(180^0)\approx -\frac{\sqrt{2}}{4}\,.
\end{equation}
As can be seen in Fig. 6, this relation is generally fulfilled and
only small deviations appear due to pion rescattering and
Fermi motion.

Thus, we find that the maximum value of the tensor analyzing
power $T_{20}$ is only sensitive  to the $D$-state component of the deuteron 
wave function and not to the details of the elementary amplitudes or to
pion rescattering.

The behavior of the other tensor polarization observables
$T_{21}$ and $T_{22}$ is similar
(see Fig. 7 and Fig. 8). They both have a maximum negative value around
$\theta_{\pi}=90^0$. As in the case of $T_{20}$ this value is determined 
by the deuteron $D$-state. Using Eq. (24), this dependence can be expressed
approximately as
\begin{mathletters}
\begin{equation}
T_{21}(\sim 90^0)\approx-\sqrt{6}\sin{2\Theta}\frac{x}{1+3\,x^2\,
\sin^2{2\Theta}}\,,
\end{equation}
\begin{equation}
T_{22}(\sim 90^0)\approx-\sqrt{\frac{3}{2}}(1-\cos{2\Theta})
\frac{x}{1+3\,x^2\,\sin^2{2\Theta}}\,.
\end{equation}
\end{mathletters}
From this equation it follows that at the maximum $T_{21}\approx 2T_{22}$.
In the forward and backward direction these observables vanish.
At low energies the $S$-state component becomes
visible and the $D$-state contribution is negligible. In this region, $T_{21}$
can be estimated by the following expression
\begin{equation}
T_{21}^{(S)}(E_{\gamma}\sim 200 MeV)\approx-\frac{\sqrt{3}}{4}
\sin{\theta_{\pi}}\frac{Re[ f_1^* f_3]}{\mid f_1 \mid^2 + 
\frac{3}{4}\sin^2{\theta_{\pi}}
(\mid f_2 \mid^2 +\frac{2}{3}\mid f_3 \mid^2)}\,.
\end{equation}

{\it\underline{Vector analyzing power}.} In our study of polarization 
observables in elastic pion-deuteron scattering (see Part I) we found
that the vector analyzing power $ iT_{11}$ is very sensitive to the details 
of the theory. Especially pion rescattering was very important, 
around the  $\Delta$-resonance region it changed the sign of $iT_{11}$.

Our calculations, shown in Fig. 9,
indicate that the influence of pion rescattering is not
important for $ iT_{11}$ in the case of $\pi^0$ photoproduction
(compare solid and dashed curves). The contribution of the
deuteron $D$-state is also small. Therefore, in this reaction the vector
analyzing power is neither sensitive to details of the rescattering
mechanism nor to nuclear structure.
Therefore, it can be expressed entirely through the elementary amplitudes
$f_1,...,f_4$. If we again neglect the small
contribution from the $f_4$ amplitude
$iT_{11}$ has a simple form,
\begin{equation}
iT_{11}^{(S)}(deuteron)\approx -\frac{\sqrt{3}}{2}\sin{\theta_{\pi}}
\frac{Im\,[f_1^*(f_2-\frac{1}{2}f_3)]}{\mid f_1 \mid^2 +
\frac{3}{4}\sin^2{\theta_{\pi}}\,(\mid f_2 \mid^2 +
\frac{2}{3}\mid f_3 \mid^2)}\,.      
\end{equation}

Furthermore, due to the small nonresonant $E_{0+}$ multipole, 
$f_1$ reduces to 
\begin{equation}
f_1\approx \cos{\theta_{\pi}}\,(3E_{1+}+M_{1+}-M_{1-})\,.
\end{equation}
Therefore, $iT_{11}$ has a $\sin{\theta_{\pi}}\cos{\theta_{\pi}}$
angular dependence and changes the sign around $\theta_{\pi}=90^0$.
Note that this result is in strong disagreement with the predictions
of Garcilazo and Guerra~\cite{Garc95}. Their vector analyzing power is
positive in the $\Delta$-resonance region at all angles.
This can occur only through differences
 in the elementary amplitude. At the same time their
predictions for the photon asymmetry and tensor analyzing powers
are very similar to ours.

Another interesting consequence which comes from Eqs. (34) and (35) is 
related to the contribution of the $E_{1+}$ multipole.
It is of a particular interest because of its resonant part 
due to the $E2$ transition in the $\gamma N\Delta$ vertex. 
The size of this quadrupole component is very sensitive to the tensor 
force in the quark-quark interaction.
Thus, this transition
could give us a measure for the deformation of the delta and,
subsequently, of the nucleon in the framework of quark models.

Fig. 10 illustrates the sensitivity of the vector analyzing power 
to the contribution of the $E2$ transition in $\gamma N\Delta$ vertex
at $\theta_{\pi}=30^0$. As follows from Fig. 9,  the
influence of other theoretical uncertainties is minimal in this region.
 However, the
presence of the $E_{1+}^{\Delta}$ multipole dramatically changes the
value of $iT_{11}$.   

Note that a similar situation also occurs in $\pi^0$
photoproduction on the nucleon. Again, if we neglect the small contribution
from the $f_4$ amplitude we obtain for the target asymmetry on the proton
\begin{equation}
T=\sqrt{2}iT_{11}(nucleon)\approx -\sin{\theta_{\pi}}
\frac{Im\,[f_1^*(f_2-f_3)]}{\mid f_1 \mid^2 +
\frac{1}{2}\sin^2{\theta_{\pi}}\,(\mid f_2 \mid^2 +
\mid f_3 \mid^2)}\,.
\end{equation}

\section{ETA PHOTOPRODUCTION}

We complete our analysis with an investigation of coherent eta 
photoproduction on the deuteron. The formalism for this process can be 
developed 
in a straightforward way by the same coupled channels method which has been 
applied to pion photoproduction. In momentum space the nuclear
photoproduction amplitude can be written as 
\begin{eqnarray}
F_{\eta \gamma}^{(\lambda)}(\vec{q},\vec{k}) = 
V_{\eta \gamma}^{(\lambda)}(\vec{q},\vec{k}) -
\frac{a}{(2\pi)^2} \sum_{{i=\pi,\eta}} \int \frac{d^3q'}{{\cal M}_i(q')} 
\frac{F_{\eta i}(\vec{q},\vec{q}\,')\,V_{i\gamma}^{(\lambda)}
(\vec{q}\,', \vec{k})}
{{\cal E}_{\eta}(q) -{\cal E}_i(q') + i \epsilon} \, \, ,
\end{eqnarray}
where $\vec{k}$ is the photon, and $\vec{q}$ is the eta or pion momentum. The 
total energy in the $\eta$-nucleus and $\pi$-nucleus channels is denoted by 
${\cal E}_i(q) = E_i(q) + E_A(q)$, the reduced mass is given by
${\cal M}_i(q) = E_i(q) E_A(q)/{\cal E}_i(q)$ and $a=(A-1)/A$.  

$V_{\eta \gamma}$ is expressed in terms of the free eta--nucleon 
photoproduction t--matrix
\begin{eqnarray}
V_{\eta \gamma}^{(\lambda)}(\vec{q},\vec{k})=
-\frac{\sqrt{{\cal M}_{\eta}(q){\cal M}_{\gamma}(k)}}{2\pi}<\eta(\vec{q}),f \mid
\sum^A_{j=1} t_{\eta\gamma}(j)\mid i,\gamma (\vec{k}\lambda)>\,.
\end{eqnarray}

As an elementary $t_{\eta\gamma}$ amplitude we will use the extended 
dynamical model of Bennhold and Tanabe~\cite{BT,TBK} where the 
resonance sector is constrained by solving the coupled channels 
problem for the $\pi N\rightarrow\pi N$, $\pi N\rightarrow\eta N$, 
$\pi N\rightarrow\pi\pi N$ and $\gamma N\rightarrow\pi N$ reactions, 
using available data as input. The background is formed by the 
contributions from $s$- and $u$-channel Born terms and by
$\rho$- and $\omega$- exchange in the $t$-channel. Following
the analysis of Ref.~\cite{TBK} we use
pseudoscalar Born terms with an $\eta NN$ coupling constant of
$g_{\eta}^2/4\pi=0.4$.

At present, little is known about the nature of the eta-nucleus 
interaction which enters in the amplitude $F_{\eta i}(\vec{q},\vec{q}\,')$.
For the elementary $\pi N\rightarrow\eta N$ process the experimental data is 
much less complete and accurate in contrast to $\pi N$-scattering data. 
There are only few theoretical studies of this reaction
\cite{BT,Bha,Oset} based on the coupled channel isobar model for the 
$\pi N,\,\, \eta N$ and $\pi\pi N$ systems. 
More detailed information about the structure of the eta-nucleus interaction 
can be found in our previous work~\cite{KTB93}. 
We only note that, just as in pion photoproduction, the eta scattering
amplitude $F_{\eta i}$ is constructed as a solution of the
Lippmann-Schwinger equation using the KMT version of multiple scattering 
theory. At present our calculations have been carried out in DWIA 
without coupling to the pion channel 
because our elementary pion photoproduction amplitude is not well 
appropriate for the photon energies $E_{\gamma}>500$ MeV where higher
nucleon resonances become important. 
From the other hand as it was estimated in Ref.\cite{Eta2}  
contributions coming from the pion channel are much smaller than 
eta rescattering terms.

In Fig. 11, we show the differential cross section for coherent eta
photoproduction on the deuteron at $E_\gamma=675 MeV$ (48 MeV above
threshold). The reaction is completely dominated by the resonance 
contribution, the non-resonant Born terms and vector meson contributions 
are negligible. The eta rescattering contribution in the final state is 
also small in this energy region.

Fig. 12 presents the energy dependence of the differential
cross section at $\theta=90^0$. Old experimental
data~\cite{Oldeta} are in dramatic disagreement
with numerous theoretical predictions~\cite{Eta1,Eta2,Eta3}.
This is related to the small isoscalar
$(\gamma,\eta)$ amplitude $E_{0+}^{(0)}$ 
with $R=E_{0+}^{(0)}/E_{0+}^{(p)}=0.22$. 
As can be seen in Fig. 12, only an unrealistically large isoscalar 
amplitude of $E_{0+}^{(0)}/E_{0+}^{(p)}=0.7$ can explain the data.  This
puzzle was a subject of investigations by many authors over the last twenty
years. Recently, a new experiment at Mainz by Krusche et al. \cite{Kru95}
has obtained an upper limit for coherent $\eta$ photoproduction
which is in agreement with our calculations using a realistic 
ratio of $R\approx 0.2$.
This is also confirmed by a preliminary analysis of an 
experiment at Bonn \cite{Ant95}.

\section{CONCLUSION}
In this paper we have presented a calculation for coherent pion and 
eta photoproduction from the deuteron using an approach that is consistent
with the description of pion elastic scattering (Part I) \cite{KTB95}.
The deuteron wave function used in our computation is obtained from the 
Paris potential. The final state interaction of the emitted mesons is 
treated in multiple scattering as described in Part I. All calculations are 
performed in momentum space treating  dependence of the elementary amplitude 
from mesons and nucleon momenta exactly. 

Our calculations agree well with existing data of differential cross 
sections and can also reproduce previous calculations found in the 
literature. In the case of $\pi^0$ photoproduction in   
the threshold region we have found significant sensitivity of
the differential cross section to the $E_{0+}^{(+)}$ isospin amplitude
of the nucleon that should allow a determination of the $E_{0+}(\pi^0 n)$
amplitude\cite{blok94}. For coherent $\eta$ photoproduction new 
experimental data from Mainz is in agreement with our results
obtained using a realistic ratio of $E_{0+}^{(0)}/E_{0+}^{(p)}=0.22$. 

We have applied our approach to polarization observables that can 
be measured with the new electron scattering facilities where polarized 
targets and beams are becoming available. A number of simple relationships 
between polarization observables and the basic production amplitude 
was derived for the kinematical regions where nuclear 
structure and pion rescattering are not important. 
We found a strong sensitivity to the deuteron $D$-state
component in the tensor analyzing powers. The vector analyzing power  
($iT_{11}$) or target polarization shows an enhanced
sensitivity to the resonant $E_{1+}^{\Delta}$ multipole. This very important
quantity may give insight into the medium modification of baryons in the 
presence of other nucleons. A study of this amplitude is possible by 
looking at 
different nuclei. In a previous paper on $^3$He it was found that this 
multipole can be rather easily observed with a polarized photon beam.

Up to now there are only few good data available for this rather elementary 
reaction and polarization observables are missing in the low energy
and $\Delta$ region. With the new 
accelerators at Mainz, Bonn, Saskatoon and, very soon, CEBAF and the 
development of very powerful spectrometers and detectors we hope that soon 
precise data will become available, thus improving our knowledge of 
the production and interaction of mesons with very light nuclei.

\acknowledgements
We would like to thank Henk Blok, Bernd Krusche and Michael Fuchs for a 
fruitful discussion on threshold photoproduction.
This work was supported by the Deutsche Forschungsgemeinschaft (SFB 201), 
the U.S. DOE grant DE-FG02-95-ER40907 and the Heisenberg-Landau program.

\appendix
\section*{}

In this appendix we give PWIA expressions for the {\it Robson
amplitudes} ${\cal A},...{\cal E}$ with the deuteron $S$- and $D$-states
taken into account. They are
expressed with nuclear matrix elements $M_{SLJ}(Q)$ defined in 
Ref.~\cite{KTB95} 
\begin{equation}
M_{000}(Q)=2\,\sqrt{\frac{3}{4\pi}}\left[R_{00}^{(0)}(Q)+
R_{22}^{(0)}(Q)\right]\,,
\end{equation}
\begin{equation}
M_{022}(Q)=4\,\sqrt{\frac{3}{4\pi}}\left[R_{02}^{(2)}(Q)-
\frac{\sqrt{2}}{4}R_{22}^{(2)}(Q)\right]\,,
\end{equation}
\begin{equation}
M_{101}(Q)=2\,\sqrt{\frac{6}{4\pi}}\left[R_{00}^{(0)}(Q)-
\frac{1}{2}R_{22}^{(0)}(Q)\right]\,,
\end{equation}
\begin{equation}
M_{121}(Q)=-2\,\sqrt{\frac{6}{4\pi}}\left[R_{02}^{(2)}(Q)+
\frac{1}{\sqrt{2}}R_{22}^{(2)}(Q)\right]\,.
\end{equation}

For the magnetic part of the ${\cal A},...,{\cal E}$ amplitudes 
in the PWIA approach we obtain
\begin{equation}
{\cal A}_M=\sqrt{\frac{4\pi}{6}}f_2\sin{\theta_{\pi}^*}
\left[M_{000}(Q) -\frac{1}{4\sqrt{2}}(1+3\cos{2\Theta})\,
M_{022}(Q)\right]\,W_A\,,
\end{equation}
\begin{equation}
{\cal B}_M=\frac{1}{2}\sqrt{\frac{4\pi}{6}}\left[ f_1\,M_S(Q)+
\frac{3}{2}f_2\sin{\theta_{\pi}^*}\sin{2\Theta}\,M_{022}(Q)\right]\,W_A\,,
\end{equation}
\begin{equation}
{\cal C}_M= -\frac{3}{8}\sqrt{\frac{4\pi}{3}}f_2\sin{\theta_{\pi}^*}
\left(1 - \cos{2\Theta}\right)\,M_{022}(Q)\,W_A\,,
\end{equation}
\begin{equation}
{\cal D}_M= -\frac{1}{2}\sqrt{\frac{4\pi}{6}}\left[f_1\,M_S(Q)-
\frac{3}{2}f_2\sin{\theta_{\pi}^*}\sin{2\Theta}\,M_{022}(Q)\right]\,W_A\,,
\end{equation}
\begin{equation}
{\cal E}_M =\sqrt{\frac{4\pi}{6}}f_2\sin{\theta_{\pi}^*}
\left[M_{000}(Q) +\frac{1}{2\sqrt{2}}(1+3\cos{2\Theta})\,
M_{022}(Q)\right]\,W_A\,.
\end{equation}

For the electric part we obtain
\begin{equation}
{\cal A}_E= -\frac{1}{2}\sqrt{\frac{4\pi}{3}}\sin{\theta_{\pi}^*}
\left[(f_3+f_4\cos{\theta_{\pi}^*})M_S(Q)-\frac{3}{\sqrt{2}}f_Q\cos{\Theta}\,
M_{121}(Q)\right]\,W_A\,,
\end{equation}
\begin{equation}
{\cal B}_E={\cal D}_E= -\frac{1}{2}\sqrt{\frac{4\pi}{6}}
\left[(f_1+f_4\sin^2{\theta_{\pi}^*})M_S(Q)+
\frac{3}{\sqrt{2}}f_Q\sin{\theta_{\pi}^*}\sin{\Theta}\,
M_{121}(Q)\right]\,W_A\,.
\end{equation}
In the PWIA approach the electric amplitude  
${\cal C}_E=0$ which contributes only in the presence of pion rescattering. 
In Eqs. (A5-A11) we used the notation
\begin{equation}
M_S(Q)=M_{101}(Q)-\frac{1}{\sqrt{2}}M_{121}(Q)\,.
\end{equation}

\begin{figure}
\caption{
(a) Reduced total cross section as a function of the photon
lab. energy near threshold. The dotted curves are the PWIA results, 
the dashed and solid curves are the calculations including pion 
charge exchange
without and with elastic pion rescattering, respectively. 
The data are from Ref.~\protect\cite{Saclay}.
(b) Reduced differential cross section for the photon excess energy in the 
lab. frame above threshold, $\Delta E=E_{\gamma}-E_{\gamma}^{thr.}= 8$ MeV. 
The notation of the curves is the same as in (a).}
\end{figure}

\begin{figure}
\caption{
(a) Sensitivity of the reduced total cross section on the
isospin (+) amplitude in the direct term. From top to bottom the 
$E_{0+}^{(+)}$ s-wave amplitude has been taken as 
$-1, -0.5, 0, +0.5, +1$ in units of $10^{-3}/m_\pi$. 
All curves have been calculated with full rescattering including 
the SCE contribution. The data are from Ref.~\protect\cite{Saclay}.
(b) Reduced differential cross section for the photon excess energy in the 
lab. frame above threshold, $\Delta E=E_{\gamma}-E_{\gamma}^{thr.}= 8$ MeV. 
The notation of the curves is the same as in (a).}
\end{figure}

\begin{figure}
\caption{
Energy dependence of the differential cross section at 
fixed angles $\theta_{\pi}=6^0,\,90^0,\,120^0$ and 175$^0$. The dotted 
and dashed curves are the calculations only with Born+$\omega$-exchange and
$\Delta$ contributions, respectively. Solid curves are the results of the
full calculations. Experimental data are from Refs.~\protect\cite{Hilg75} (o), 
~\protect\cite{Holt73} ($\bullet$) and \protect\cite{Beul94} ($\triangle$).}
\end{figure}

\begin{figure}
\caption{
Angular distribution at fixed photon lab. energy $E_{\gamma}$=
260, 300, 340 and 400 MeV. The dashed and solid curves are PWIA and
full calculations. The dash-dotted and dotted curves are the calculations 
without non spin-flip and without deuteron $D$-state, respectively. 
Experimental data are from Refs.~\protect\cite{Hilg75} (o), 
Refs.~\protect\cite{Holt73} ($\bullet$) 
and \protect\cite{Bouq74} ($\triangle$).}
\end{figure}

\begin{figure}
\caption{
Photon asymmetry $\Sigma$ at $E_{\gamma}$=200--400 MeV. 
The solid and dashed curves are full and PWIA calculations, respectively. 
The dotted curves are obtained without deuteron $D$-state. }
\end{figure}

\begin{figure}
\caption{
Tensor analyzing power $T_{20}$. The notations of the curves are
as in Fig. 5.}
\end{figure}

\begin{figure}
\caption{
Tensor analyzing power $T_{21}$. The notations of the curves are
as in Fig. 5.}
\end{figure}

\begin{figure}
\caption{
Tensor analyzing power $T_{22}$. The notations of the curves are
as in Fig. 5.}
\end{figure}

\begin{figure}
\caption{
Vector analyzing power $T_{11}$. The notations of the curves are
as in Fig. 5.}
\end{figure}

\begin{figure}
\caption{
Sensitivity of $iT_{11}$ to the $E_{1+}^{\Delta}$ multipole.
The solid and dashed curves are the results obtained with and without 
this multipole, respectively.}
\end{figure}

\begin{figure}
\caption{
Differential cross section for coherent eta photoproduction on the 
deuteron at 675 MeV photon lab. energy. The dotted curves show the 
individual contributions for the resonance excitation and the non-resonant
background from vector mesons and Born terms in PWIA. The dashed and full 
lines give the full calculations in PWIA and DWIA, respectively.}
\end{figure}

\begin{figure}
\caption{
Energy dependence of the differential cross section at 
fixed angles $\theta_{\eta}=90^0$ for the $d(\gamma,\eta)d$ reaction 
calculated with different ratios $R=E_{0+}^{(0)}/E_{0+}^{(p)}=0.22,\,0.5,\,$ 
and 0.7,  where the first value is the prediction of the Bennhold Tanabe 
model~\protect\cite{BT}. The dashed and solid curves are the PWIA and DWIA 
calculations, respectively. Experimental data are from 
Ref.~\protect\cite{Kru95} (o) and \protect\cite{Oldeta} ($\bullet$).}
\end{figure}

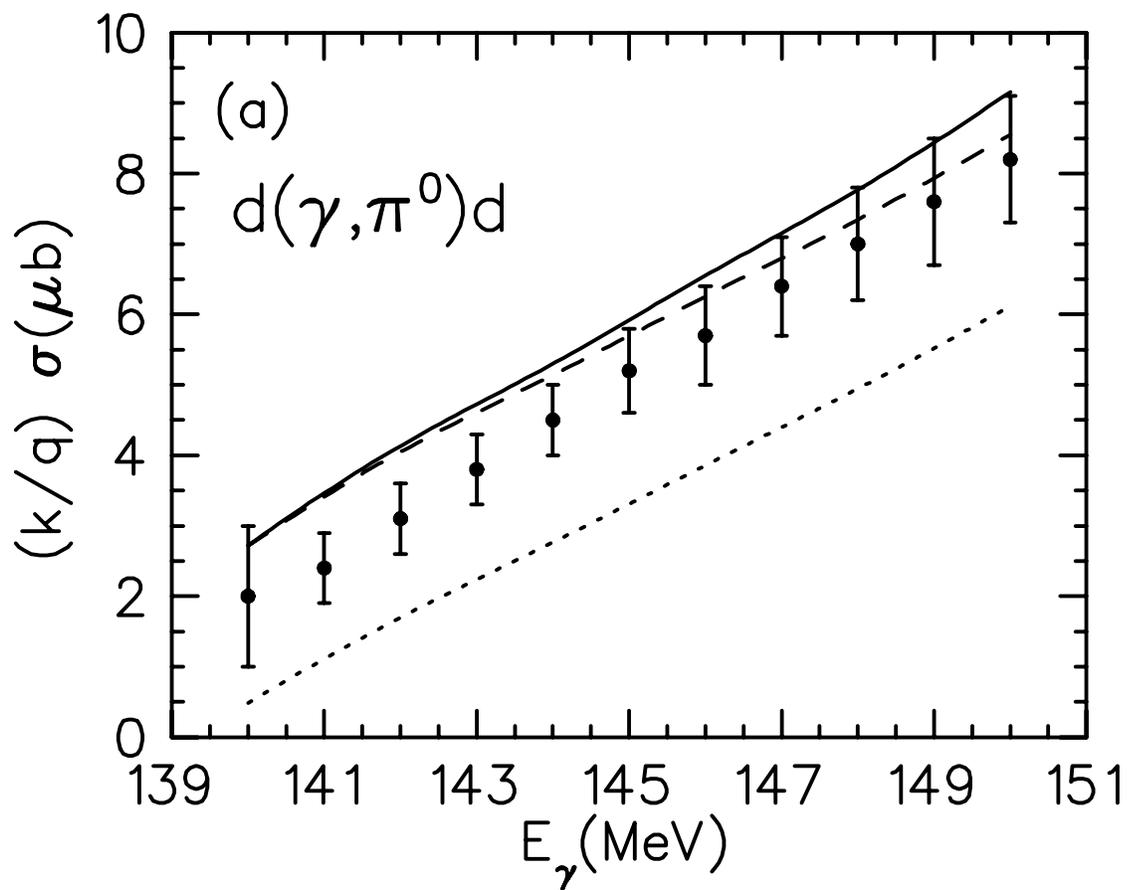
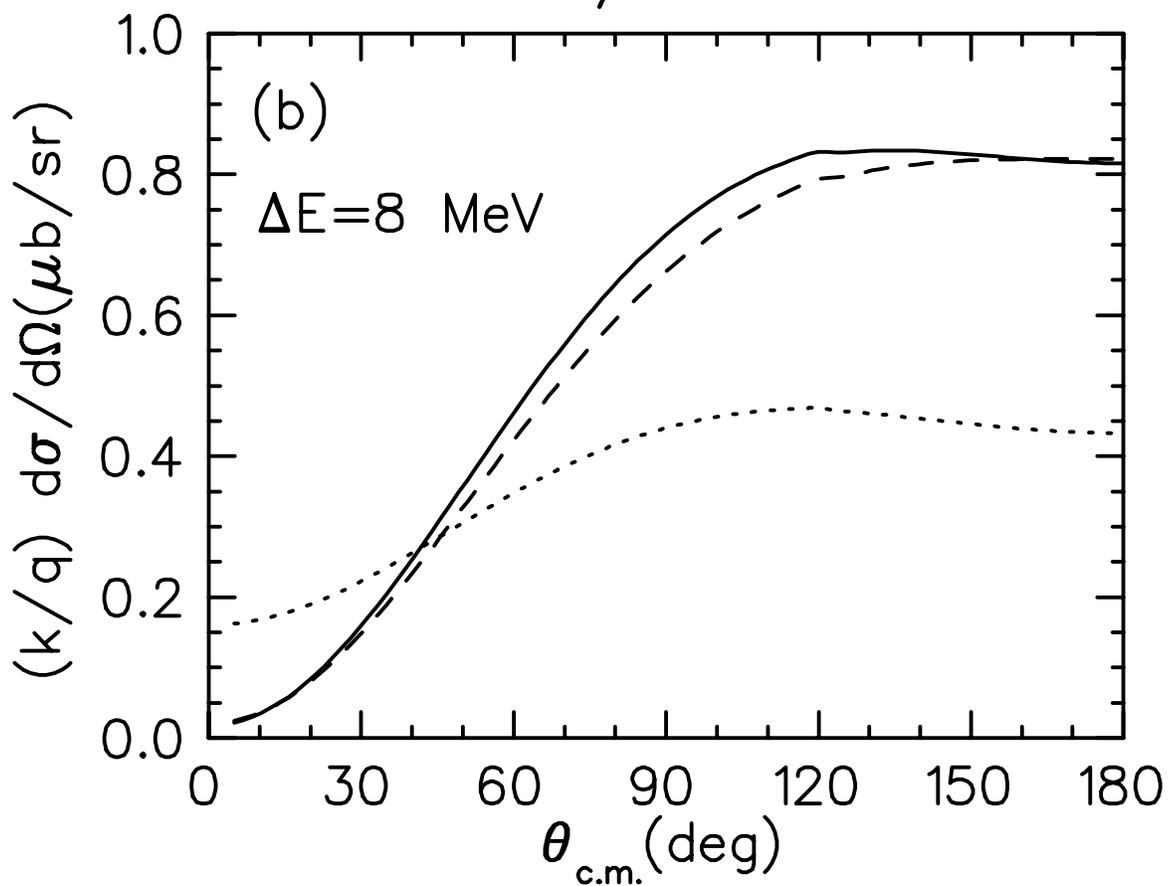
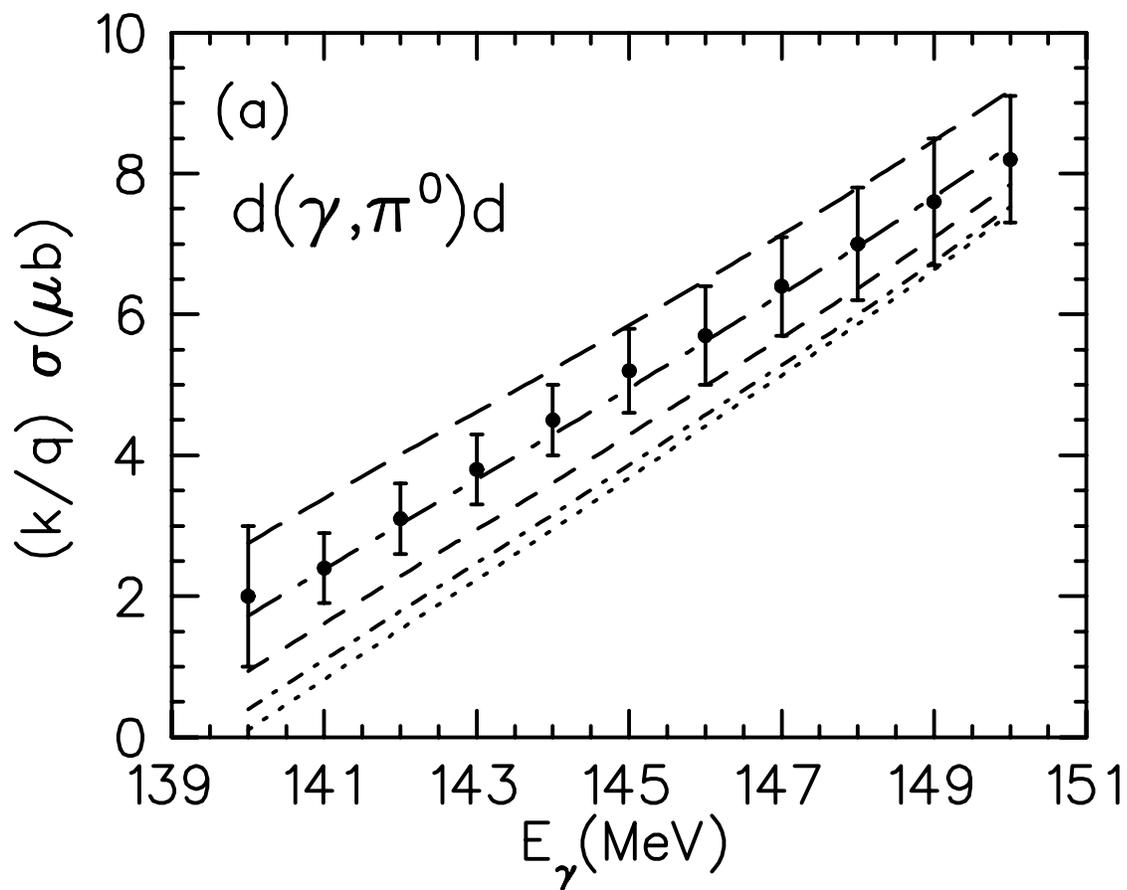
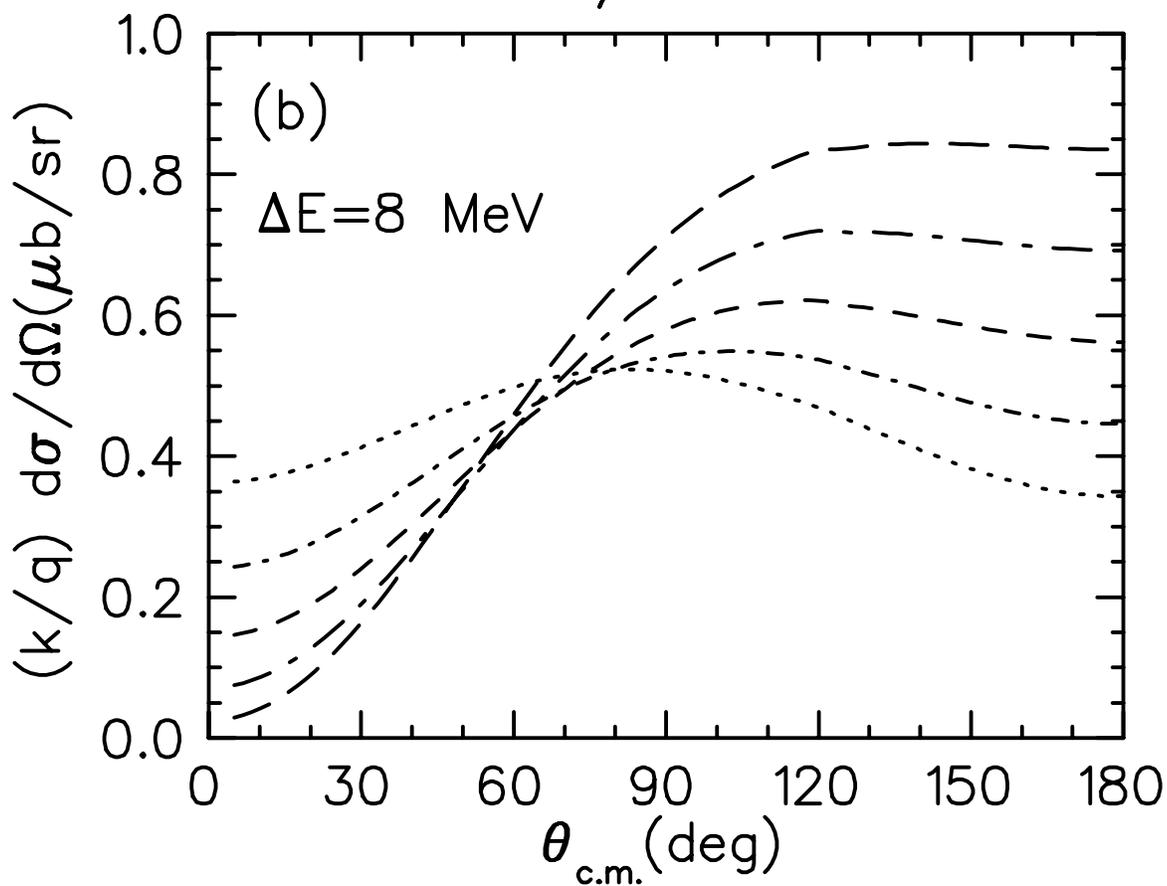
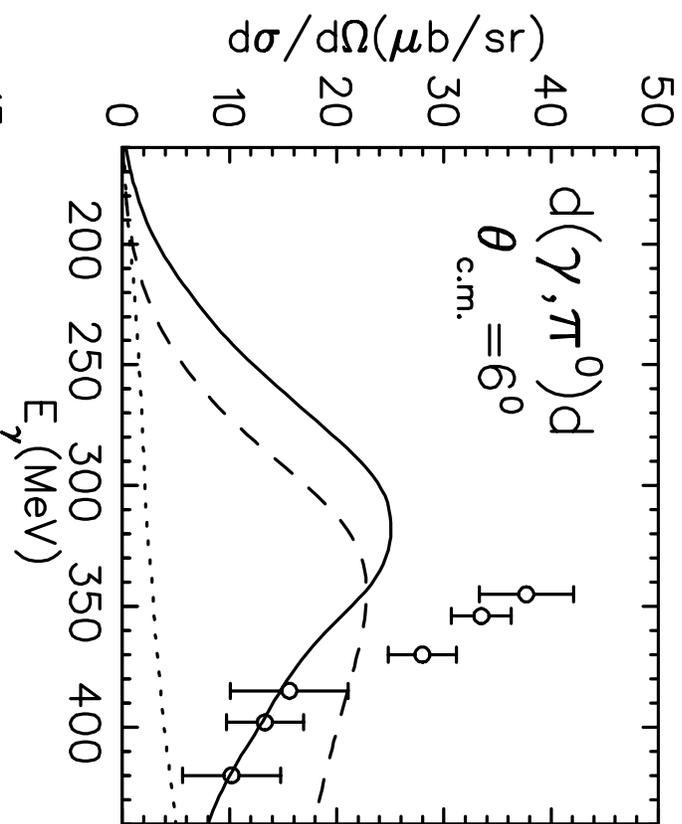
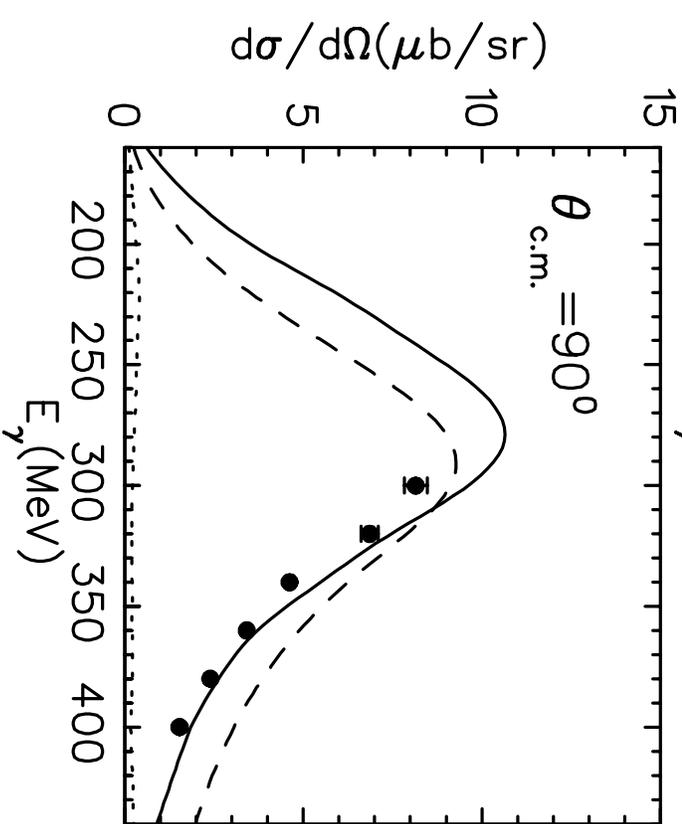
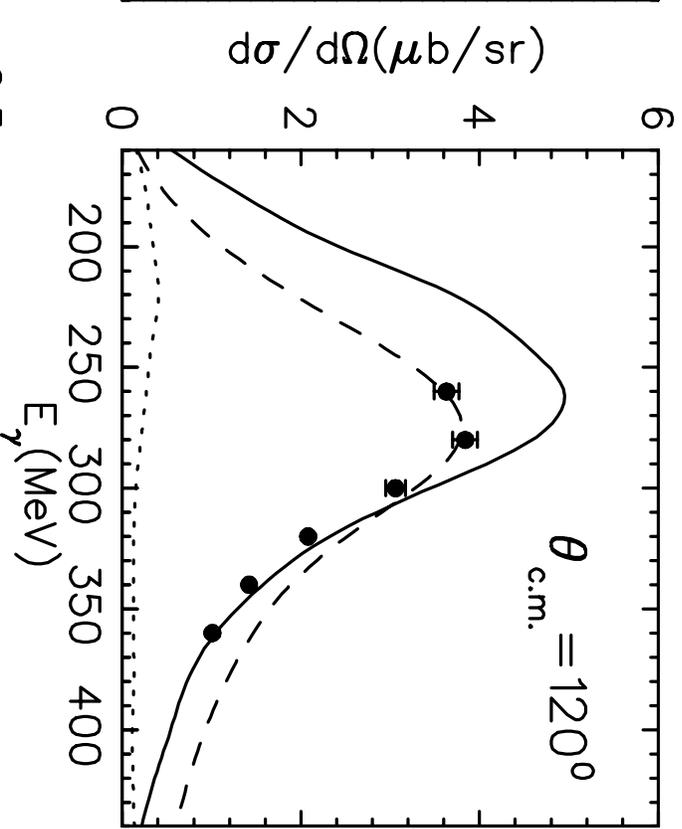
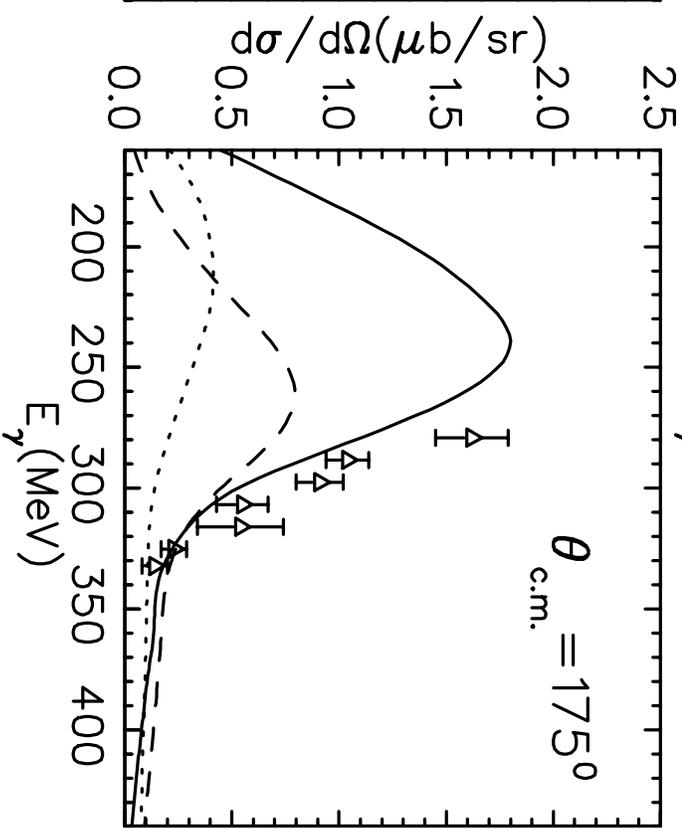
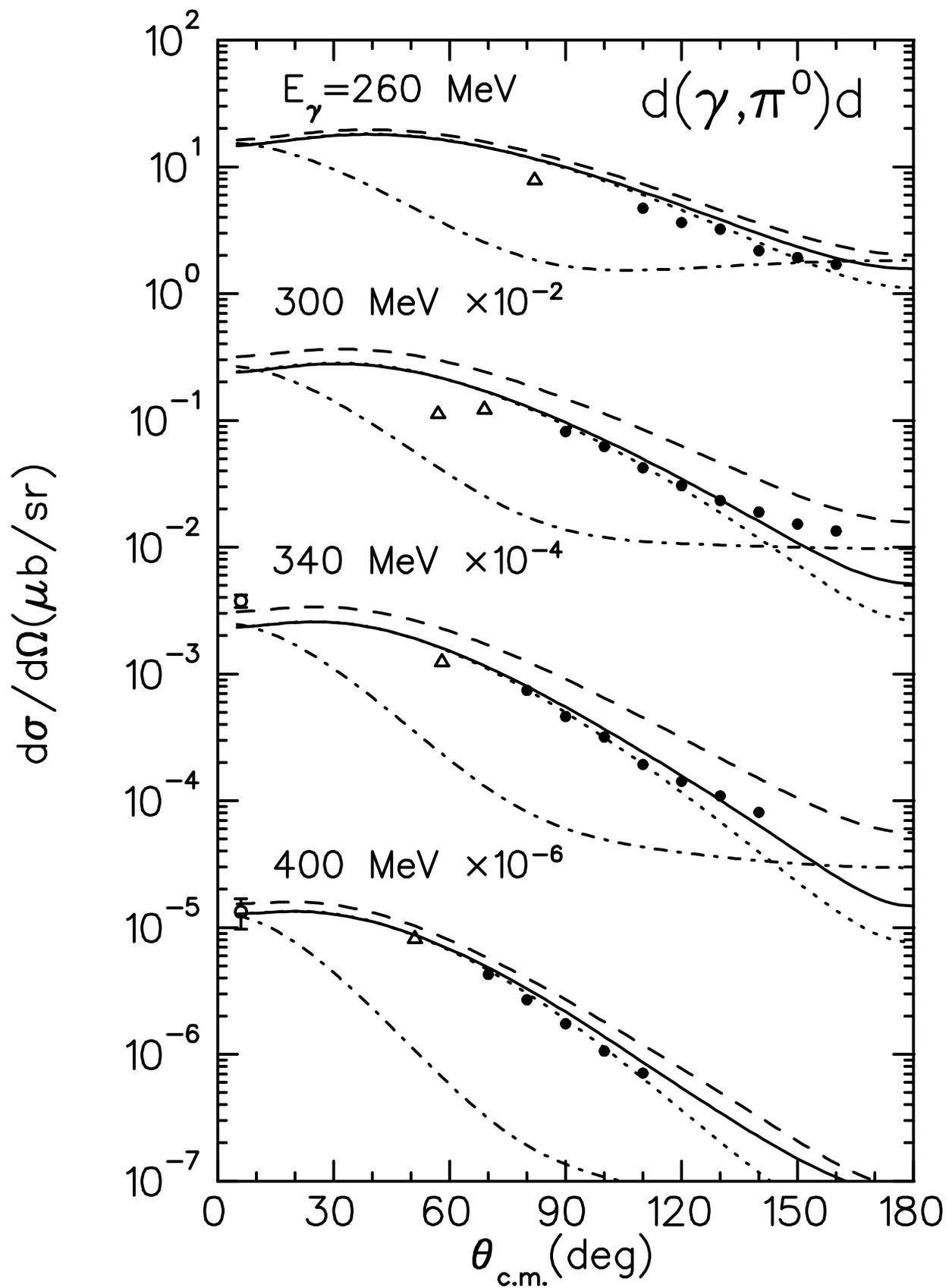
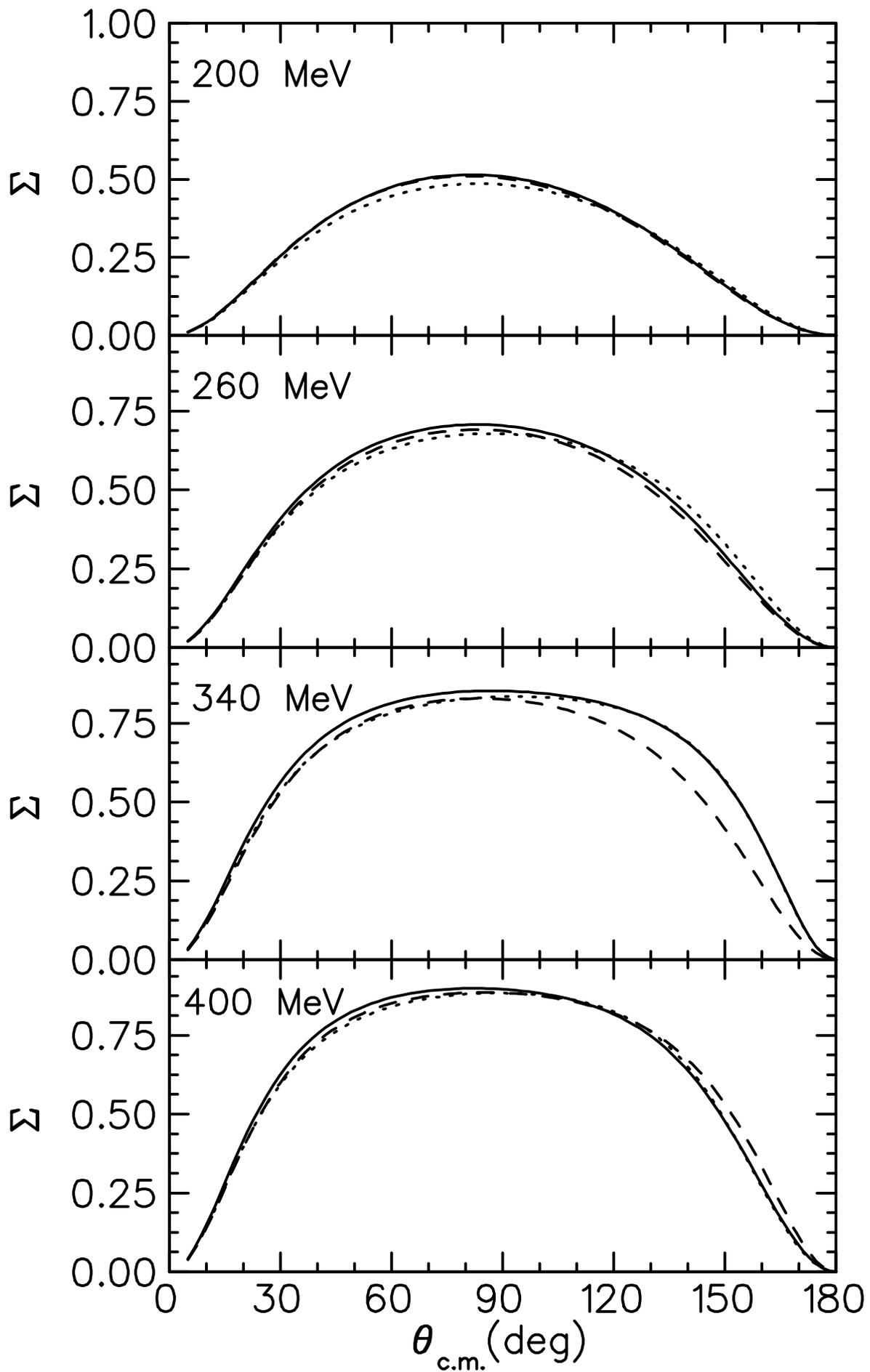
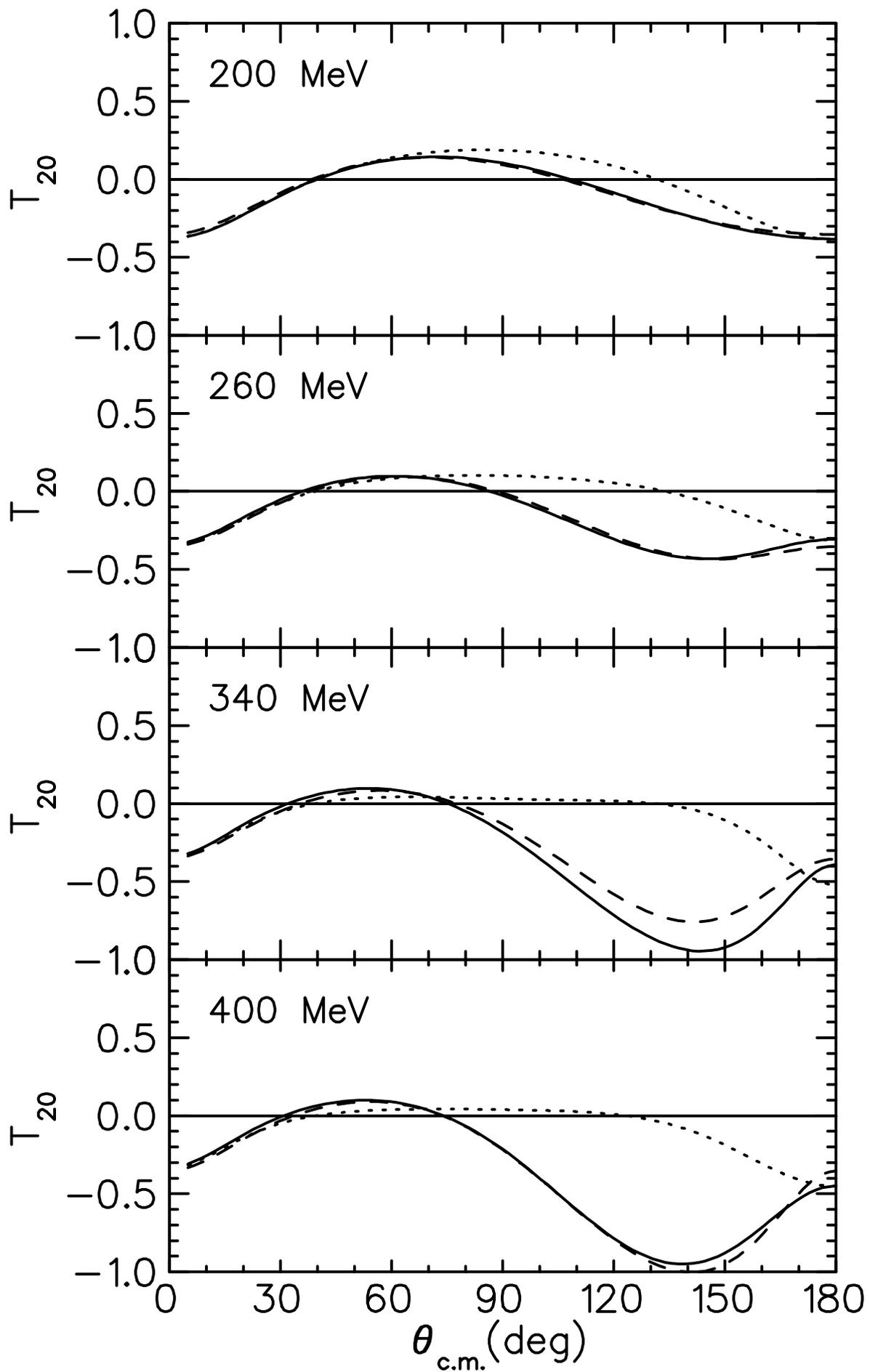
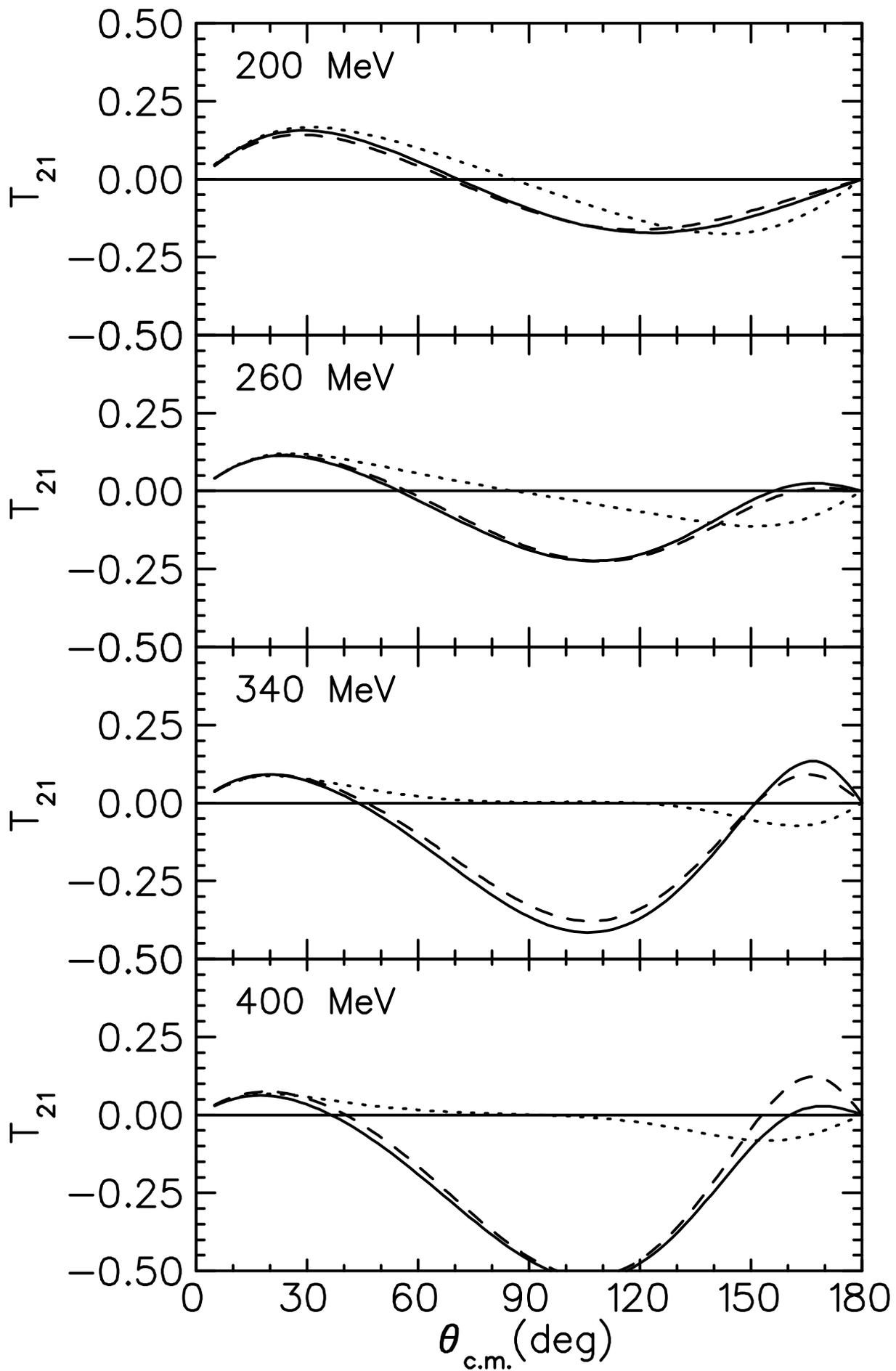
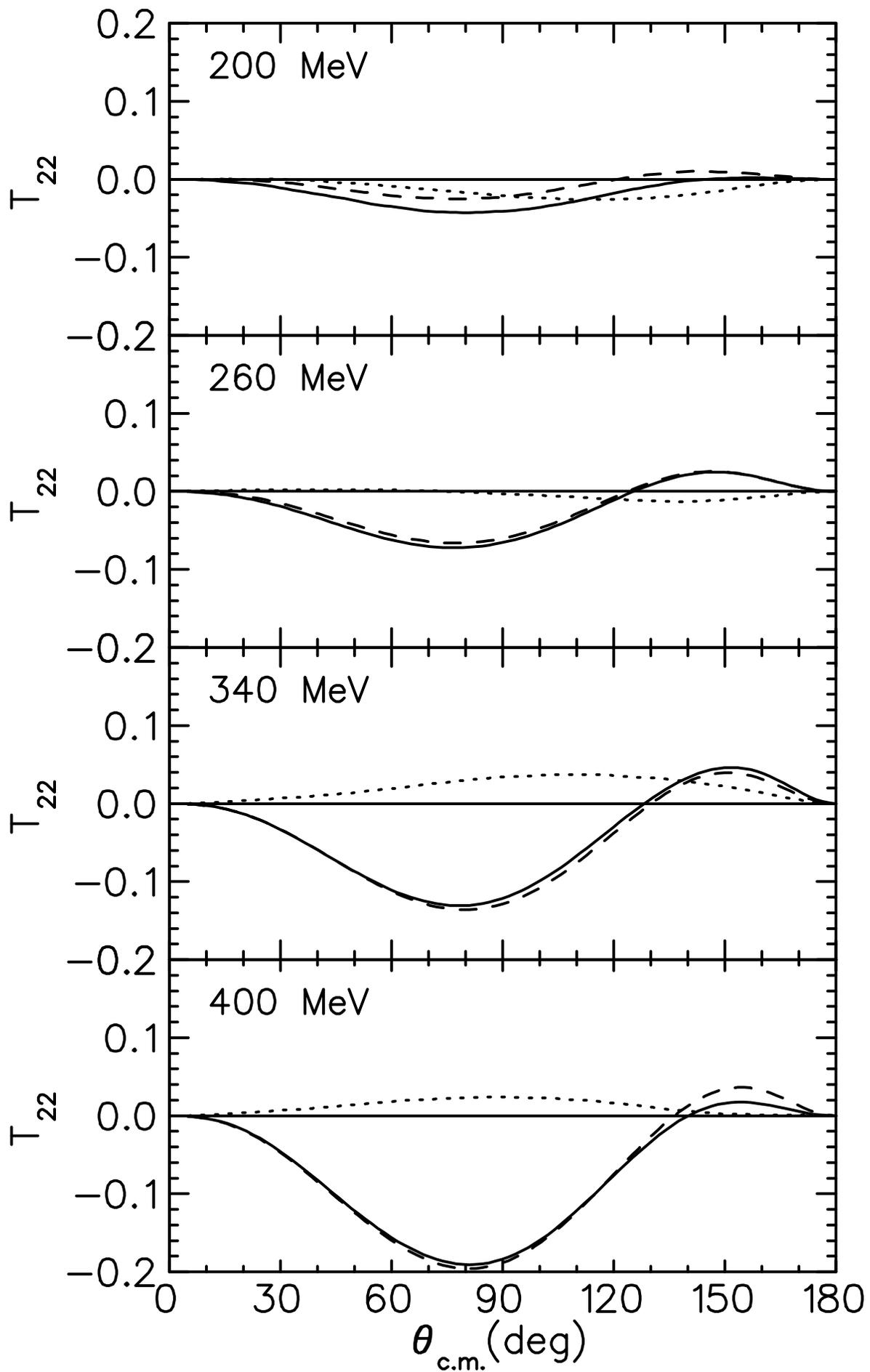
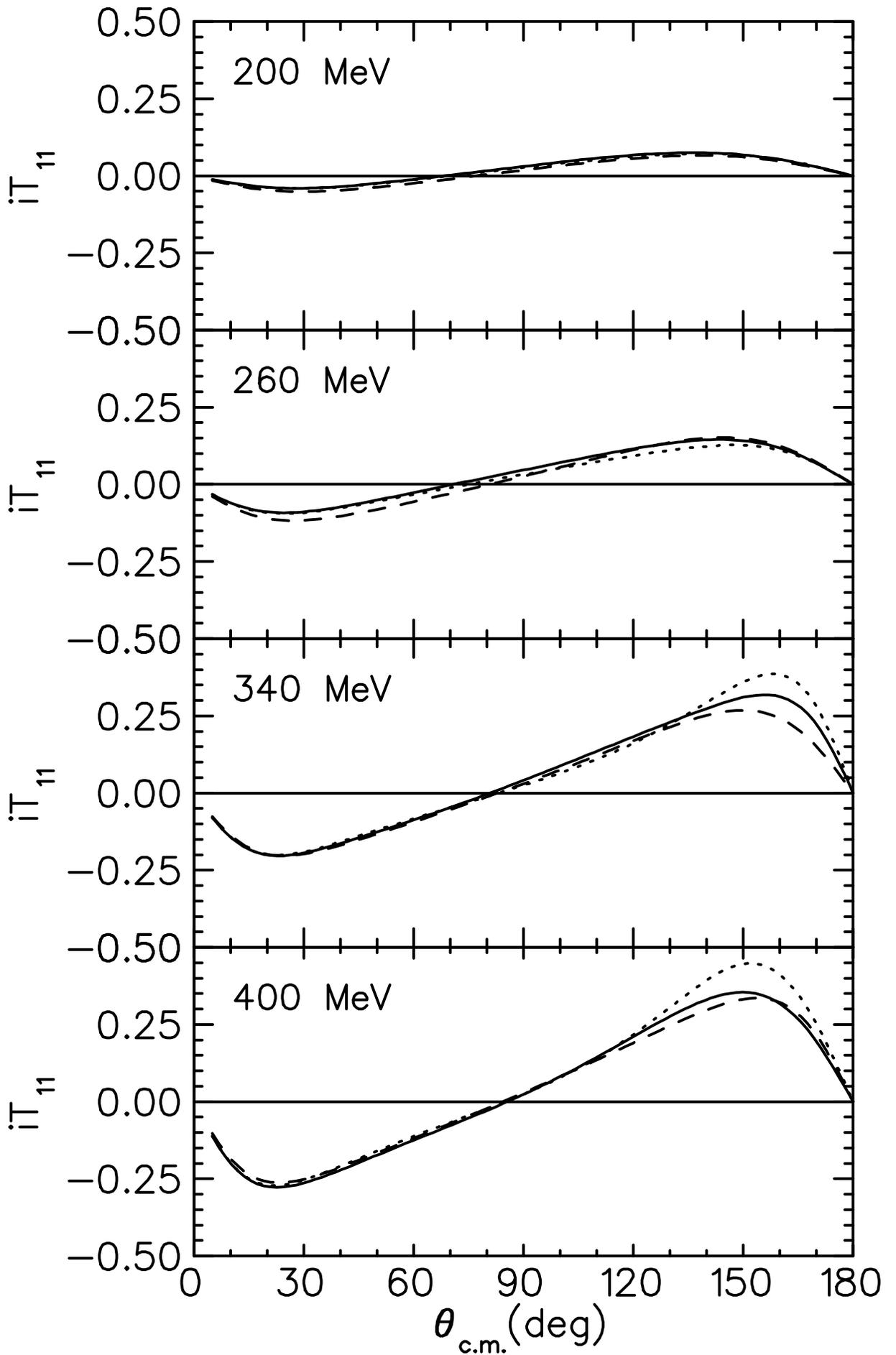
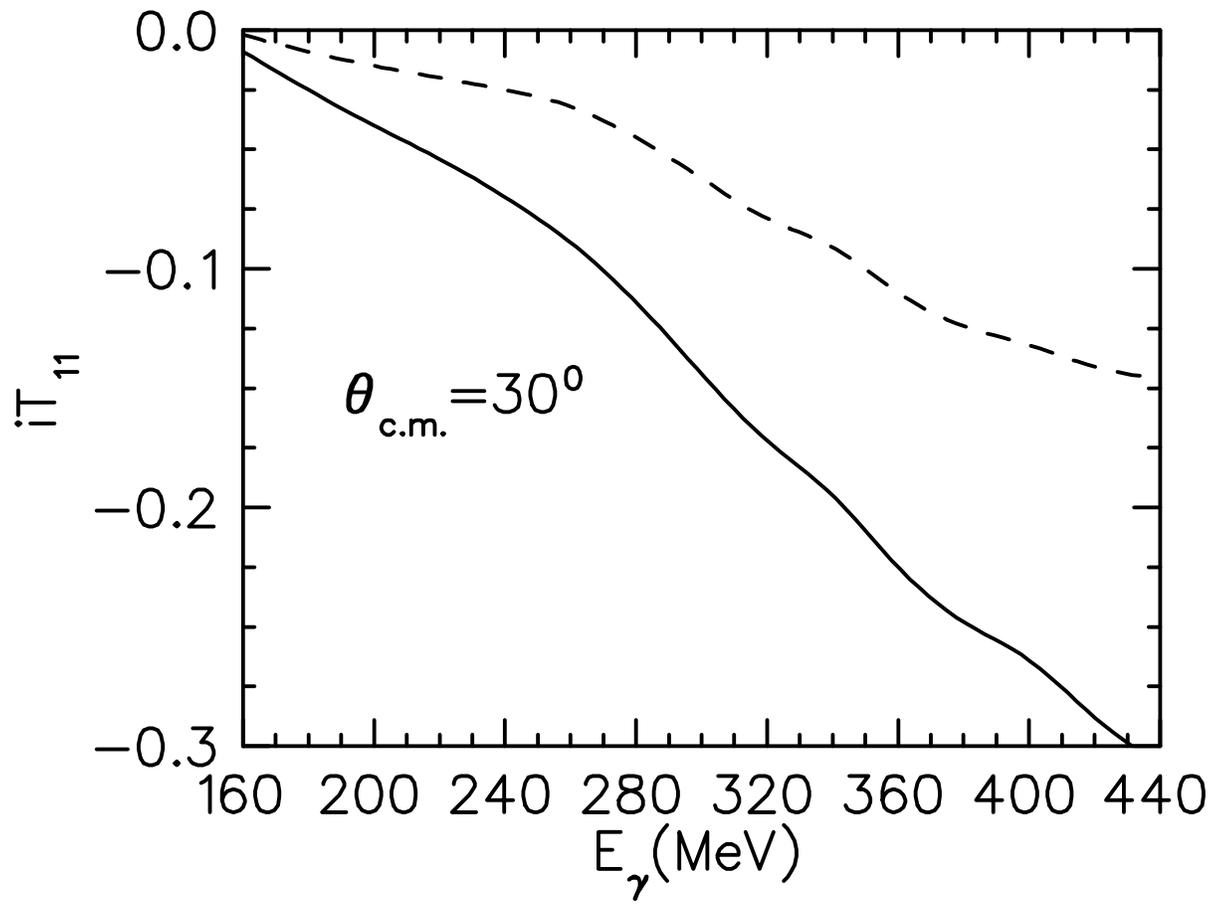
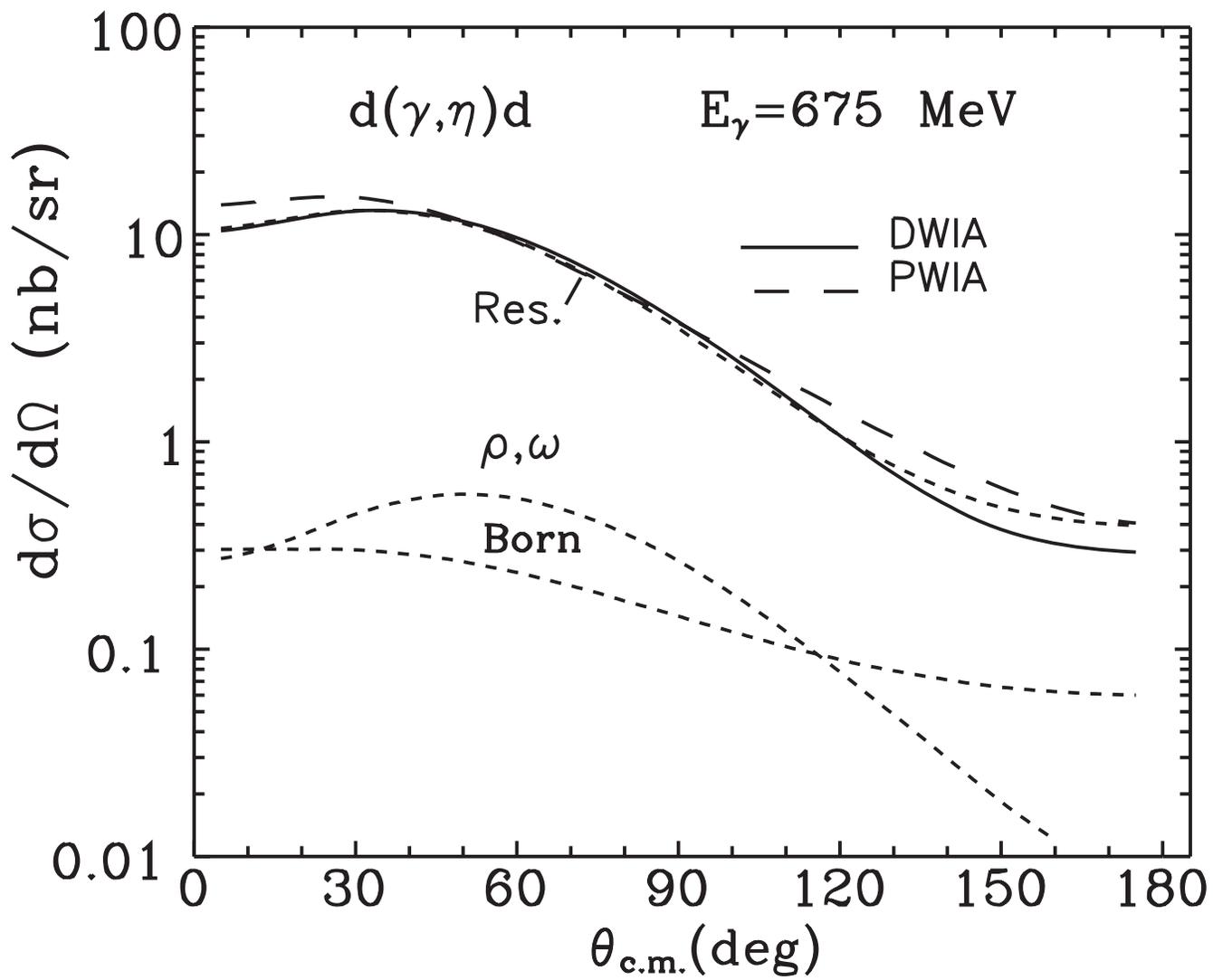
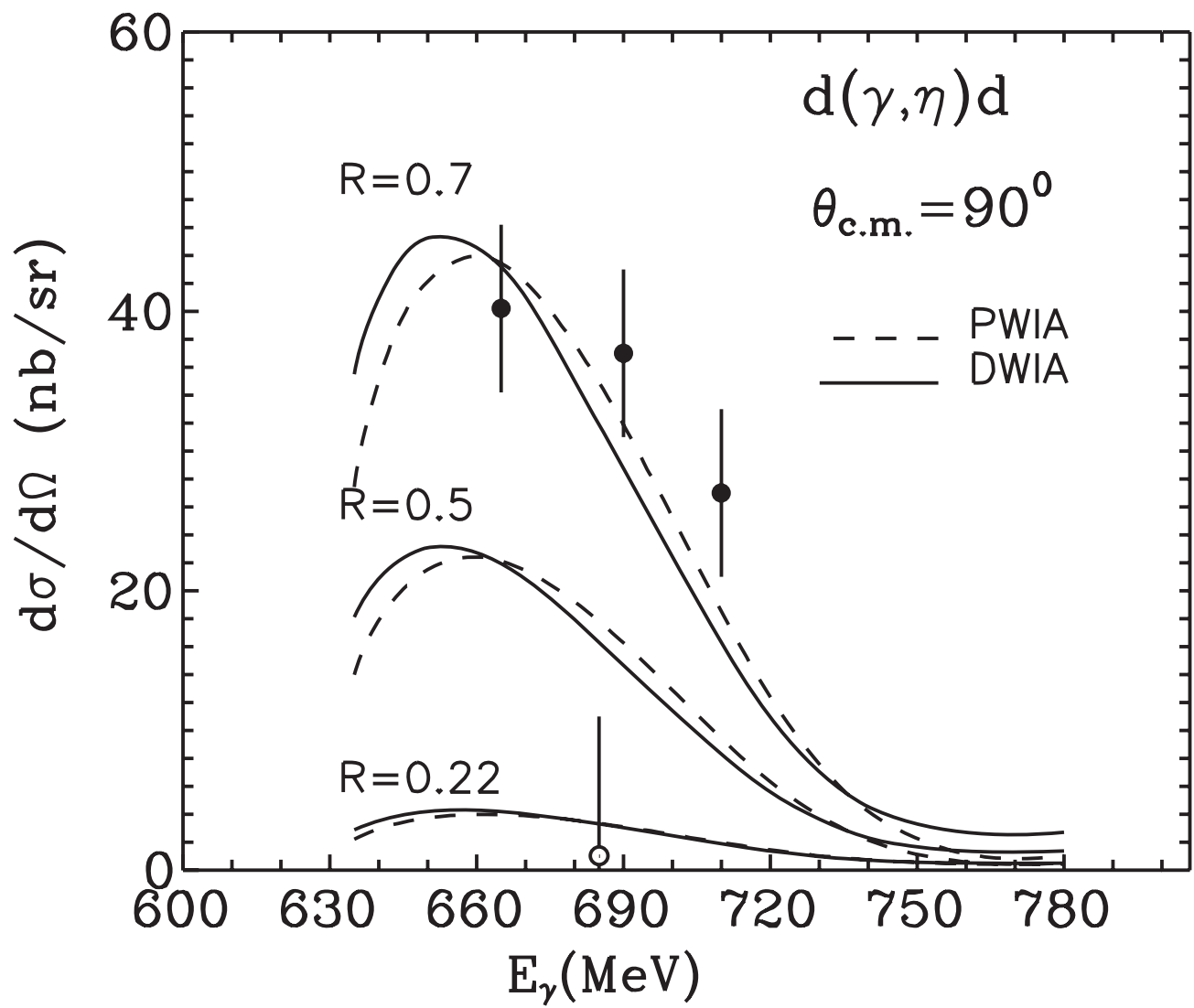
\end{document}